\documentclass[aip,amsmath,amssymb,preprint]{revtex4-1}

\usepackage{graphicx}
\begin{document}


\title{Mixed cationic clusters of nitrogen and hydrogen} 



\author{P. Martini}%
\author{F. Hechenberger}%
\author{M. Goulart}%
\author{J. Zelger}%
\author{P. Scheier}%

\affiliation{Institut f\"{u}r Ionenphysik und Angewandte Physik, Universit\"{a}t Innsbruck, Technikerstr.~25, A-6020 Innsbruck, Austria}%

\author{M. Gatchell}
   \email{michael.gatchell@uibk.ac.at}
 \affiliation{Institut f\"{u}r Ionenphysik und Angewandte Physik, Universit\"{a}t Innsbruck, Technikerstr.~25, A-6020 Innsbruck, Austria}

 \affiliation{Department of Physics, Stockholm University, 106 91 Stockholm.}


\date{\today}

\begin{abstract}
The addition of small impurities, such as a single proton charge carrier, in noble gas clusters has recently been shown to have considerable effects on their geometries and stabilities. Here we report on a mass spectrometric study of cationic clusters of N$_2$ molecules and the effects that adding hydrogen, in the form of D$_2$, has on the systems. Protonated nitrogen clusters formed by the breakup of D$_2$ are shown to have similar behaviors as protonated rare gas clusters. For larger systems consisting of different mixtures of intact N$_2$ and D$_2$, the different molecular species are found to sometimes be interchangeable with regards to magic numbers. This is especially true for the (N$_2$)$_n$(D$_2$)$_m$D$^+$ systems with $n+m=17$, which is particularly abundant for all measured combinations of $n$ and $m$. 

\end{abstract}


\maketitle 

\section{Introduction}

Nitrogen compounds have attracted interest in recent years as a potential source of clean energy.\cite{Ma:2017aa,Li:2018aa} The triple bond in N$_2$ is particularly stable and one of the strongest molecular bonds available\cite{Xu:2016aa} (9.8\,eV), and reactions increasing the bond orders between N atoms could potentially be used to release significant amounts of stored chemical energy. And while the number of all-nitrogen molecular species vastly dwarfs that of, for example, all-carbon molecules, a large amount of work has gone into identifying such species and their properties, including the theoretical prediction of a fullerene-like N$_{60}$.\cite{Samartzis:2006aa}

The smallest molecule of pure nitrogen, the ubiquitous N$_2$, has been the target of numerous studies in the context of cluster physics. Cationic clusters of N$_2$ have been produced and studied by means of different experimental techniques, such as ion impact on nitrogen solids at cryogenic temperatures,\cite{Jonkman:1981aa,Tonuma:1994aa,Fernandez-Lima:2007aa,Ponciano:2008aa} by electron impact\cite{Friedman:1983aa,Scheier:1988ab,Leisner:1988aa,Scheier:1988aa,Mark:1989aa,Walder:1991aa} or photoionization\cite{Flesch:2004aa} of gas phase clusters. These have identified even-numbered cluster series consisting of intact N$_2$ molecules as well as products resulting from the fragmentation of such a molecule, giving clusters with odd numbers of N atoms. The latter in particular have attracted the attention of theory for describing the reaction products formed.\cite{Nguyen:2001aa,Li:2002aa,Evangelisti:2003aa,Law:2002aa} Nitrogen molecules have also found use as weakly interacting molecular tags in ion spectroscopy applications.\cite{Nowak:1988aa,Baer:2010aa,Craig:2017aa}

Negatively charged clusters of N$_2$ have also been produced in doped helium nanodroplets\cite{Weinberger:2017aa} and in supersonic expansion,\cite{Vostrikov:2006aa} and were later studied theoretically.\cite{Yurtsever:2019aa} These experiments showed that the odd-numbered (N$_2$)$_n$N$_3^-$ were the dominant species, with magic sizes appearing at $n=4$ and 11,\cite{Weinberger:2017aa} and it was predicted that these were formed by the packing of N$_2$ molecules around the stable azide core. Theory showed that this was indeed a plausible explanation, with the molecules were arranged parallel to the core.\cite{Yurtsever:2019aa} However, they also showed that simple interaction potentials could not fully describe the interactions as the charge will delocalize over the surrounding molecules.\cite{Yurtsever:2019aa}

Weakly interacting clusters of molecules share many characteristics with atomic rare gas clusters,\cite{Friedman:1983aa,Calvo:1999aa} which have attracted a significant amount of interest since the early 1980s. Countless studies of rare gas clusters since then have shown such systems form highly symmetric structures where abundance anomalies in the measured mass spectra can be explained by simple sphere packing models and the closure of structural (sub-)shells.\cite{Echt:1981aa,Friedman:1983aa,Harris:1986aa,Scheier:1987aa,Mark:1987aa} More recently it was shown that these simple models, that work well for neutral systems, can break down for charged clusters, and that protonated rare gas clusters generally have structures that better match these models than the pure cationic systems.\cite{Gatchell:2018aa,Gatchell:2019ac} In light of this, we here report on measurements of nitrogen clusters mixed with hydrogen (D$_2$), which we use to produce protonated clusters. Helium droplets doped with molecular nitrogen and deuterium have been studied in the past, in a work that focused on ion-molecule reactions and the formation of N$_2$D$^+$.\cite{Farnik:2005aa} In this present study we expand the scope to include clusters containing up to 30 N$_2$ molecules. We identify and discuss differences in the mass spectra between pure (N$_2$)$^+$ clusters and the protonated systems, as well as how the mixtures of the two species behave. Particularly noteworthy is that N$_2$ and D$_2$ molecules are found to be partially interchangeable with regards to stable structures and we argue why this is the case.

\begin{figure*}
\includegraphics[width=6in]{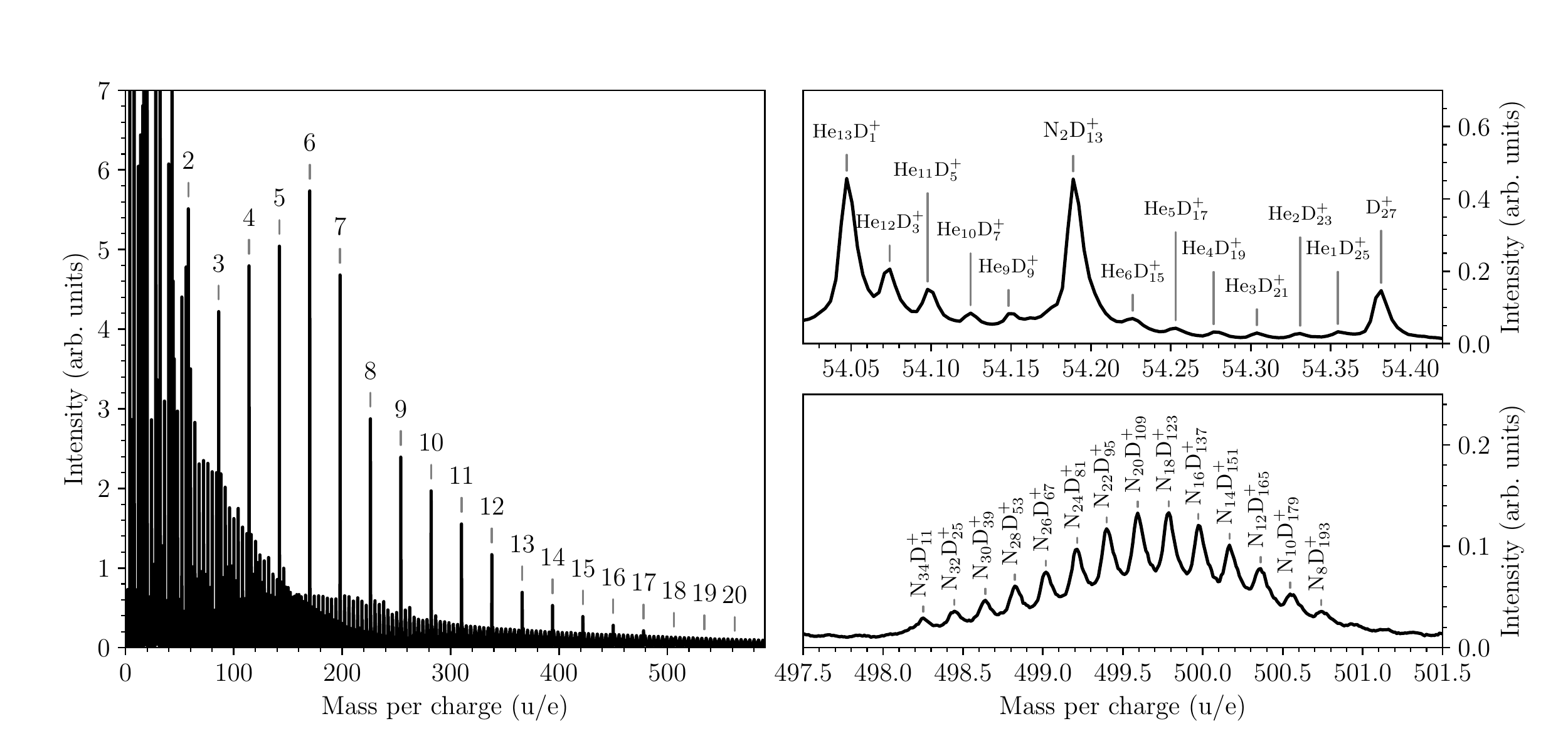}
\caption{Left: Overview mass spectrum from clusters containing N$_2$ and D$_2$ formed by electron impact on doped He nanodroplets. The dominating series that is labeled consists of (N$_2$)$_n$D$^+$ clusters. Also visible are contributions from He$_n^+$ and D$_2^+$ clusters, as well as mixes of these. Right: Two panels showing zoomed in areas illustrating the complexity and resolution of the mass spectra. In the top right panel, we can distinguish between many different mixtures of He/D clusters together with the N$_2$D$_{13}^+$ peak. In the lower right panel, several peaks from different N/D mixtures are clearly discernible.}
\label{fig:ms}
\end{figure*}

\section{Methods}
\subsection{Experimental Details}

Droplets of superfluid He were formed in the expansion of compressed and cooled He gas (Messer, 99.9999\% purity) through a 5\,$\mu$m nozzle into vacuum. The stagnation pressure of the He prior to expansion was 25.0 bar and the nozzle was cooled by a two stage closed cycle cryocooler to 10.0\,K. Under these conditions, droplets were formed with a broad log-normal size distribution where the mean droplet size is on the order of 10$^5$ He atoms.\cite{Toennies:2004aa} The central part of the He beam then passed through a 0.8\,mm skimmer positioned 8\,mm downstream from the nozzle. After the skimmer the neutral droplets traverse a pair of pickup chambers where they interacted with N$_2$, followed by D$_2$ gas, capturing molecules along the way that could condense to form cold clusters within the droplets. Deuterium is used in order to reduce overlap with clusters containing one or more $^{15}$N isotopes, but the results should be equivalent to using H$_2$. The doped droplets were subsequently ionized by the impact of electrons with kinetic energies of 60\,eV. This ionizes He atoms near the surface of the droplets, after which the holes migrate through the droplets, attracted by the higher polarizability of dopants relative to the surrounding He. In this way, multiple charge centers may be formed that may expel each other, including charged dopants from the droplets\cite{Laimer:2019aa} producing bare, charged dopant clusters. The charge transfer from the dopant to a helium cation is a highly exothermic process due to the high ionization energy of He. This will lead to a strong heating of the clusters. During the time in which they are expelled from the droplets and later fragment further in the gas phase, they will cool and particularly stable systems and structures will be enhanced in the final population. Our results thus probe the stabilities of the charged systems with little regard to the initial neutral complexes (meaning that the pickup order plays little to no role in the final products). The charged products were analyzed using a reflectron time-of-flight mass spectrometer (Tofwerk AG model HTOF) with a rated mass-resolution of 5000. The mass spectra were reduced using the IsotopeFit software,\cite{Ralser:2015aa} which allows for the extraction of clusters distributions while correcting for isotopic distributions. This method for producing and analyzing weakly bound clusters has been used on numerous occasions in the past and shown to give results that are directly comparable to other methods for cluster production, such as supersonic expansion.\cite{Gatchell:2018aa}

\subsection{Theoretical Details}

Quantum chemical structure calculations have been performed to investigate the structures of our clusters using the Gaussian 16 software.\cite{Frisch:2016aa} Based on test calculations involving comparisons with coupled cluster calculations (CCSD/cc-pVTZ) as a reference for bond lengths, angles, and charge distributions, we found that calculations using Hartee-Fock level of theory provided comparable structures for the pure, cationic N$_2$ clusters, while density functional theory (DFT) using the M06-HF functional worked well for protonated systems.

\section{Results and Discussion}
A mass spectrum of cationic products formed by the ionization of He droplets doped with N$_2$ and D$_2$ is shown in Figure \ref{fig:ms}. In the overview shown in the left panel, the dominant feature is the series of (N$_2$)$_n$D$^+$ clusters that are labeled according to their $n$-values. In addition to this series, contributions from He$_n^+$, D$_m^+$, mixtures of these, as well as different mixtures of (N$_2$)$_n$D$_m^+$ litter the spectrum. In all cases, systems with and odd number of D atoms dominate as the ionization process readily leads to the breakup of a D$_2$ molecule, producing a D$^+$ ion. In the rest of the text, the term ``protonated'' is used to describe cluster ions carrying a D$^+$ ion. The two right panels of Figure \ref{fig:ms} show detailed views of narrow mass ranges of the same spectrum, highlighting the amount of features that can bee identified in the measurements. As can be seen in the right panels, a benefit of the large mass defect of the deuterium atom is that isobaric clusters containing He or N$_2$ can be easily distinguished from systems with different numbers of D atoms.

\begin{figure}
\includegraphics[width=\columnwidth]{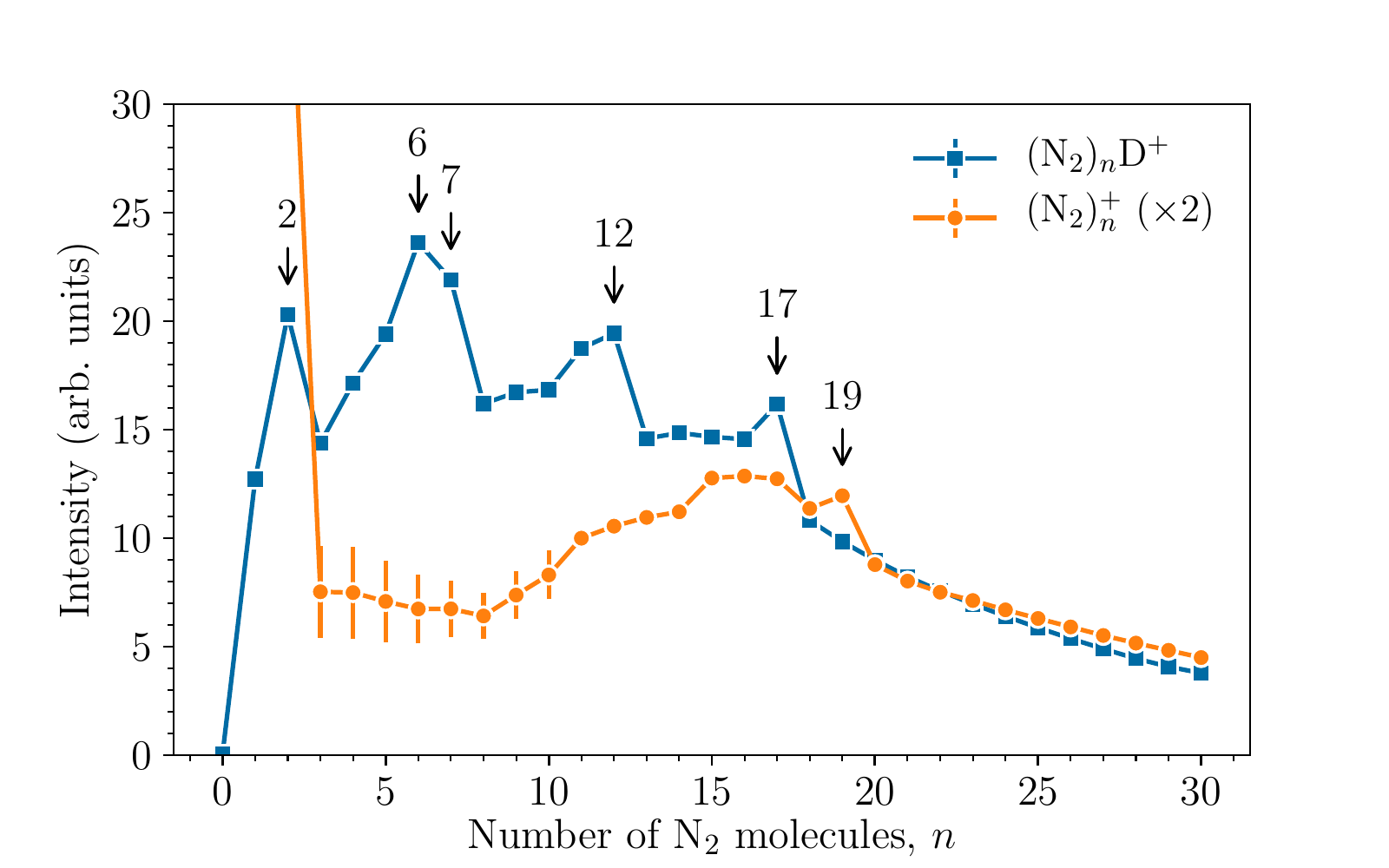}
\caption{Extracted cluster size distributions for pure cationic N$_2$ clusters (orange circles) and protonated (deuteronated) nitrogen clusters (blue squares). The protonated sequence shows a number of magic sizes, most prominently at $n=2, 6, 7, 12$, and $17$. The pure (N$_2$)$_n^+$ series is less feature rich, with $n=19$ being the only clear abundance anomaly. The larger errorbars of the latter data set are the result of the overlap with the He$_n^+$ cluster series at low masses.}
\label{fig:N2N2D_dists}
\end{figure}

The intensities of the (N$_2$)$_n$D$^+$ and (N$_2$)$_n^+$ cluster series have been extracted using the IsotopeFit software, which allows us to separate and integrate the different contributions in the mass spectrum, and are shown in Figure \ref{fig:N2N2D_dists}. For the protonated series, a number of magic sizes are clearly visible: most notably $n = 2,6,7,12$, and $17$. In contrast, the pure (N$_2$)$_n^+$ clusters have fewer distinct features, with $n=19$ being somewhat more abundant than neighboring sizes. In the pure cluster series, the error bars at small sizes are relatively large due to the overlap with He$_n^+$ clusters, which are omnipresent at low masses for clusters produced in He nanodroplets. 

The appearance of multiple distinct magic numbers for the protonated clusters mimics results with protonated rare gas clusters.\cite{Gatchell:2018aa,Gatchell:2019ac} In the case of rare gases, the presence of a proton has a stabilizing effect on the clusters compared to pure, cationic systems. This leads to magic numbers for the protonated systems that match predictions for neutral rare gases and sphere packing models.\cite{Echt:1981aa,Gatchell:2018aa} Here, the magic numbers for (N$_2$)$_n$D$^+$, most notably $n=2$, 7, 12, and 17, match the packing of N$_2$ molecules around a (N$_2$-D-N$_2$)$^+$ ionic core in icosahedral (sub-)shells, like was seen for the rare gases.\cite{Gatchell:2018aa,Gatchell:2019ac}

\begin{figure}
\includegraphics[width=0.7\columnwidth]{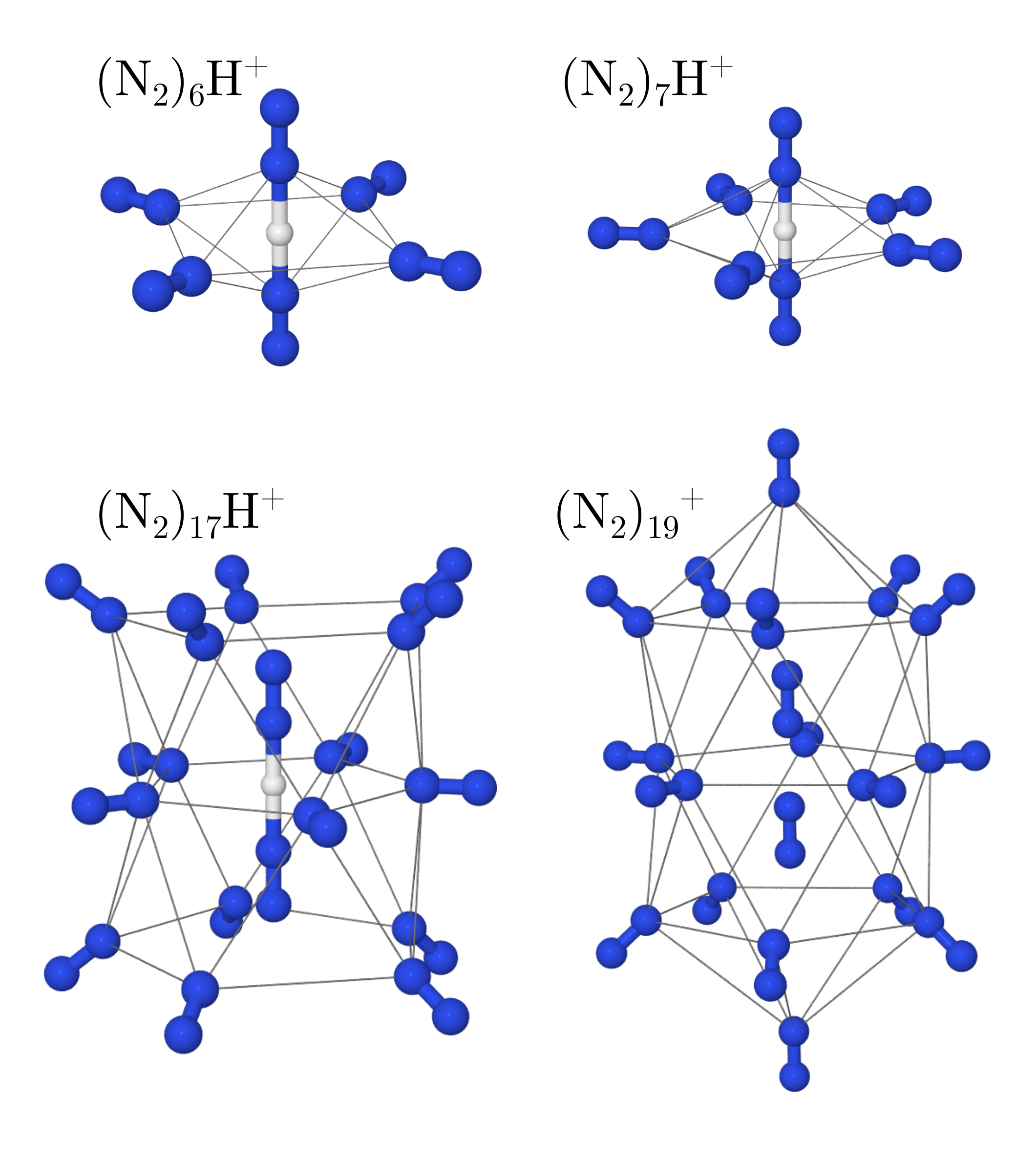}
\caption{Proposed ground state structures of (N$_2$)$_n$H$^+$ ($n=6$, 7, and 17) and (N$_2$)$_{19}^+$ clusters from quantum chemical structure calculations (M06-HF/def2-TZVPP level for protonated systems and HF/def2-TZVPP level for (N$_2$)$_{19}^+$). Thin lines are used to aid the reader in identifying the three dimensional structures and the atomic coordinates are given in the supplementary information.}
\label{fig:structs}
\end{figure}

Calculated structures of (N$_2$)$_n$H$^+$ clusters are shown in Figure \ref{fig:structs} where the packing of N$_2$ molecules around the ionic core is evident. For the bare (N$_2$-D-N$_2$)$^+$ ion, the calculated Mulliken charge distribution shows that the proton carries a net charge of 0.4\,e, with the rest of the charge being shared equally with the two nitrogen molecules. Ion-induced dipole interactions then bind additional N$_2$ molecules as the sizes of the protonated clusters increases. Our calculations indicate that the octahedral structure of (N$_2$)$_6$H$^+$ is slightly favorable over (N$_2$)$_7$H$^+$ with its pentagonal bipyrimid structure, which is consistent with the pair of magic numbers at these sizes found in the experimental data and their relative abundances. Additional N$_2$ molecules then form two more pentagonal rings above and below the initial ring to give closed sub-shells consistent with the $n=12$ and 17 magic numbers. Also shown in Figure \ref{fig:structs} is the calculated structure for (N$_2$)$_{19}^+$, the most prominent abundance anomaly in the experimental data. This structure contains a (N$_2$)$_2^+$ ionic core, where the charge is delocalized over the two center molecules which form a tightly bound system, much like as is seen for cationic rare gas clusters.\cite{Gatchell:2018aa} The structure of this cluster is similar to the (N$_2$)$_{19}$H$^+$ with an additional pair of N$_2$ molecules forming end-caps along the central axis. The shorter length of the (N$_2$)$_2^+$ ion in the pure clusters compared to (N$_2$-D-N$_2$)$^+$ in the protonated systems allows the additional molecules to favorably interact with the cluster, giving this larger magic size. Again, this matches the trends observed for protonated rare gas clusters.\cite{Gatchell:2018aa,Gatchell:2019ac}

\begin{figure}
\includegraphics[width=\columnwidth]{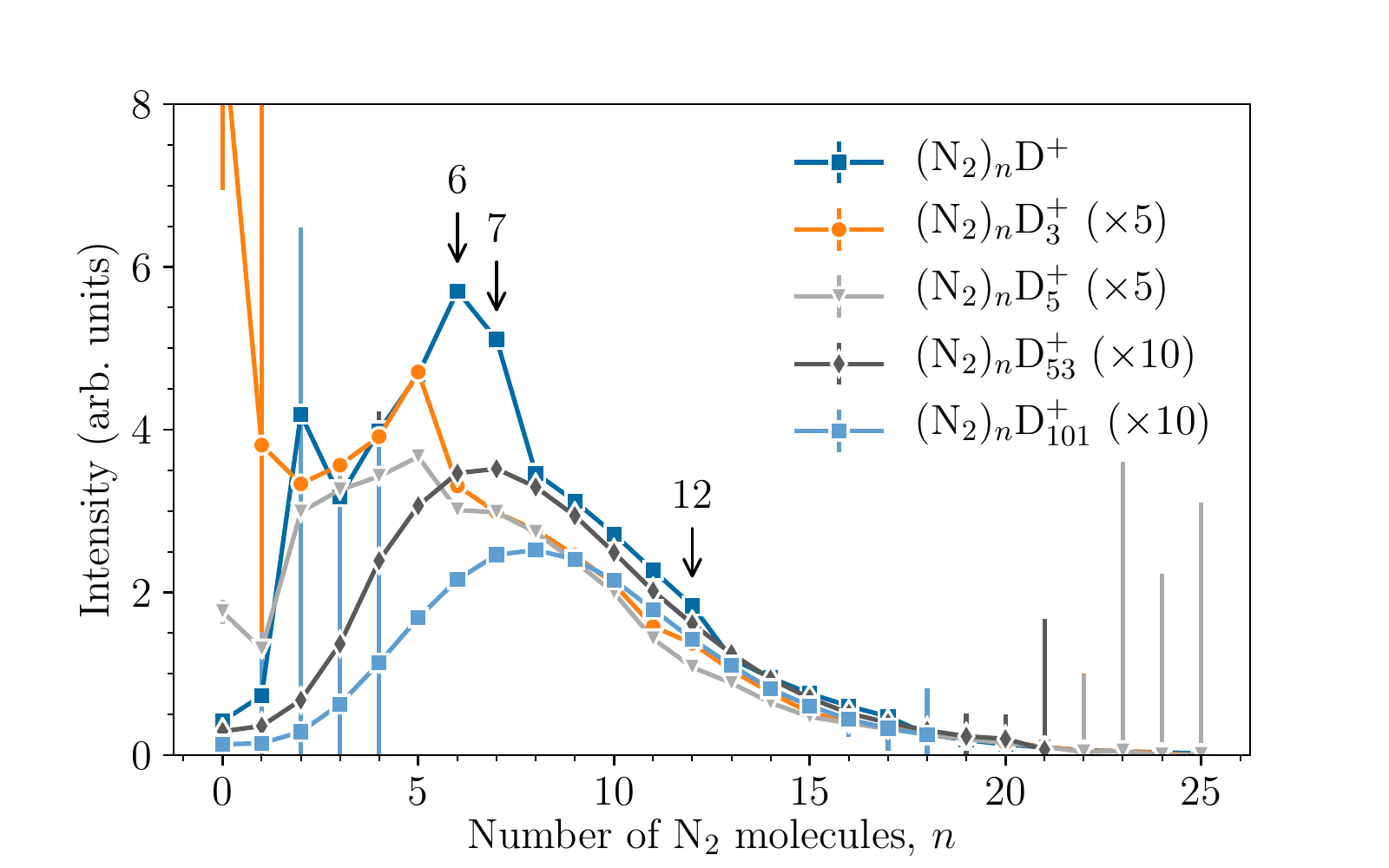}
\caption{Cluster size distributions from measurements with a larger fraction of deuterium than in the prior figure. The magic sizes $n=2, 6$, and $12$ are again visible for the (N$_2$)$_n$D$^+$ clusters. Adding D$_2$ molecules shifts the positions of magic sizes, most notably with a $n=5$ peak being dominant for both (N$_2$)$_n$D$_3^+$ and (N$_2$)$_n$D$_5^+$ clusters. An example is also shown for larger systems containing 53 and 101 D atoms, respectively. Here, there are no magic cluster sizes visible and the N$_2$ number series follow smooth distributions.}
\label{fig:N2N2Dx_dists}
\end{figure}

Increasing the relative yield of deuterium in our apparatus, we are able produce clusters mixes of N$_2$ and intact D$_2$. Figure \ref{fig:N2N2Dx_dists} shows a selection of extracted ion yields as a function of the number of N$_2$ molecules for systems containing 1, 3, 5, and 53 D atoms. Compared to the data shown in Figure \ref{fig:N2N2D_dists}, the mean N$_2$ cluster size is somewhat smaller now, but the (N$_2$)$_n$D$^+$ distribution is otherwise consistent between the measurements. The (N$_2$)$_n$D$_3^+$ distribution is similar to the (N$_2$)$_n$D$^+$ series, with the main features shifted by one N$_2$ molecule (most notably $n = 6 \rightarrow 5$), presumably due to a D$_2$ molecule replacing a N$_2$ unit in the protonated clusters. However, with the addition of another D$_2$ molecule, the $n=5$ feature is still the most prominent anomaly, indicating that the deuterium molecules are not a perfect substitute in the nitrogen clusters for preserving their structures. The last curve in Figure \ref{fig:N2N2Dx_dists} shows two examples of much larger systems, arbitrarily chosen, with 53 and 101 D atoms, respectively, together with $n$ N$_2$ molecules. These distributions are completely smooth with no distinct abundance anomalies, but there is a noticeable shift in the N$_2$ distribution towards larger sizes for the clusters with more deuterium. This is to be expected as larger parent He droplets will, on average, pick up more of both N$_2$ and D$_2$ than smaller droplets due to their larger geometrical cross sections. The smooth distributions are consistent with pickup statistics and indicates that the N$_2$ molecules are mixed with deuterium and do not form solvated clusters as this would be expected to result in features similar to the gas phase (N$_2$)$_n^+$ series in Figure \ref{fig:N2N2D_dists}.

\begin{figure}
\includegraphics[width=\columnwidth]{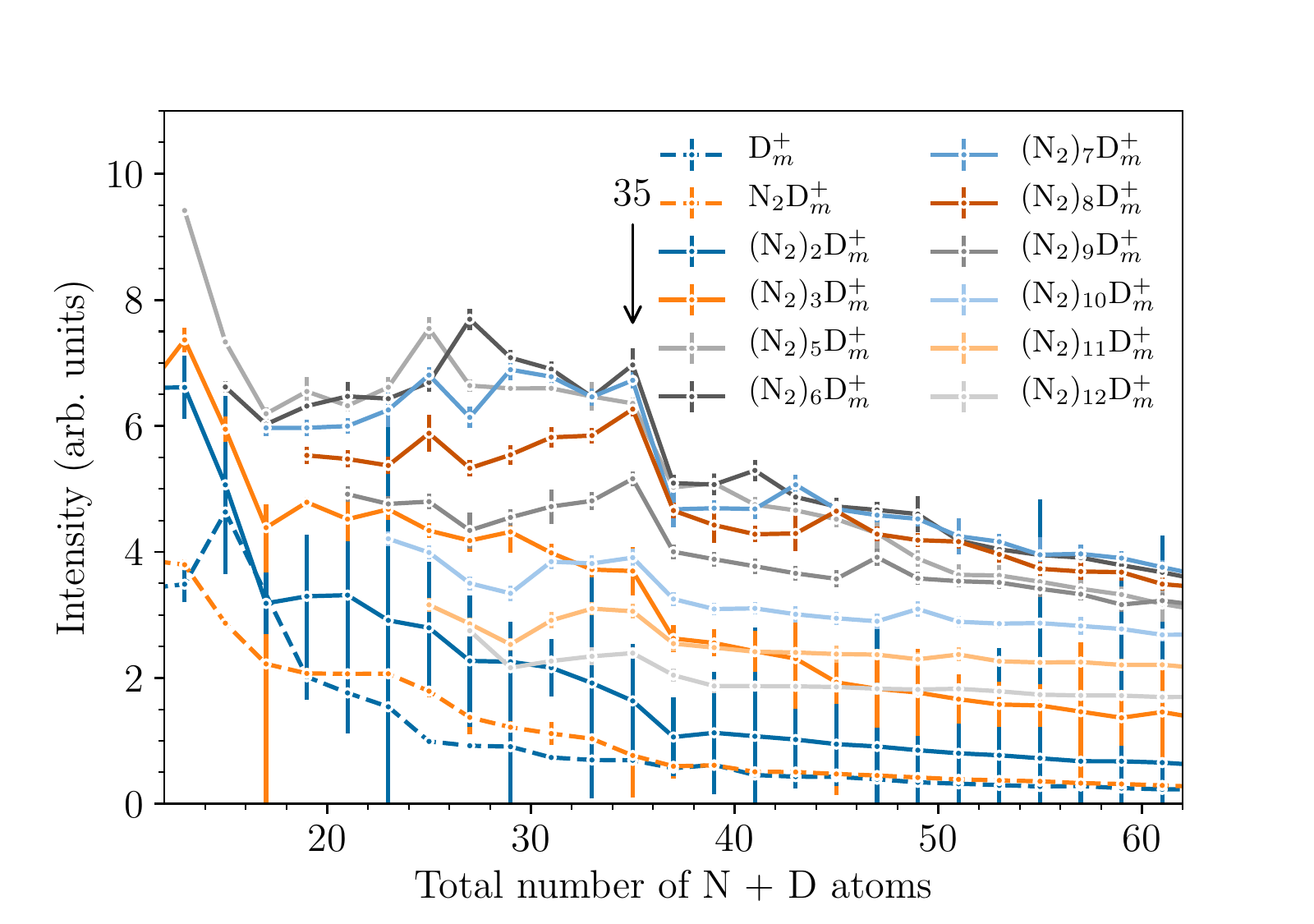}
\caption{Ion abundances for different mixtures of nitrogen and deuterium, displayed as a function of the total number of atoms. A distinctive feature of all curves is the drop in intensity after 35 atoms, corresponding to protonated systems with the total sum of N$_2$ and D$_2$ molecules equals 17.}
\label{fig:NDSums}
\end{figure}

In Figure \ref{fig:NDSums} we show distributions for several different mixtures of nitrogen and deuterium, displayed as a function of the total number of atoms in the clusters. Note that only odd numbers are shown as systems with an odd number of D atoms, i.e. closed electronic shell protonated clusters, dominate. Something that immediately stands out is a sharp drop in the intensity after 35 atoms, which is present for all measured mixtures, corresponding to (N$_2$)$_n$(D$_2$)$_m$D$^+$ clusters where $n+m=17$. Abundance anomalies are present in some series at other sizes too, but none of these are present in all mixtures containing at least two N$_2$ molecules. For pure, odd-numbered D$_n^+$ clusters, there is no distinct feature at $n=35$ and the effect is weak for clusters containing only a single N$_2$ molecule. The (N$_2$-D-N$_2$)$^+$ ion might thus form the central core for each of the different mixtures when possible.

\begin{figure}
\includegraphics[width=0.7\columnwidth]{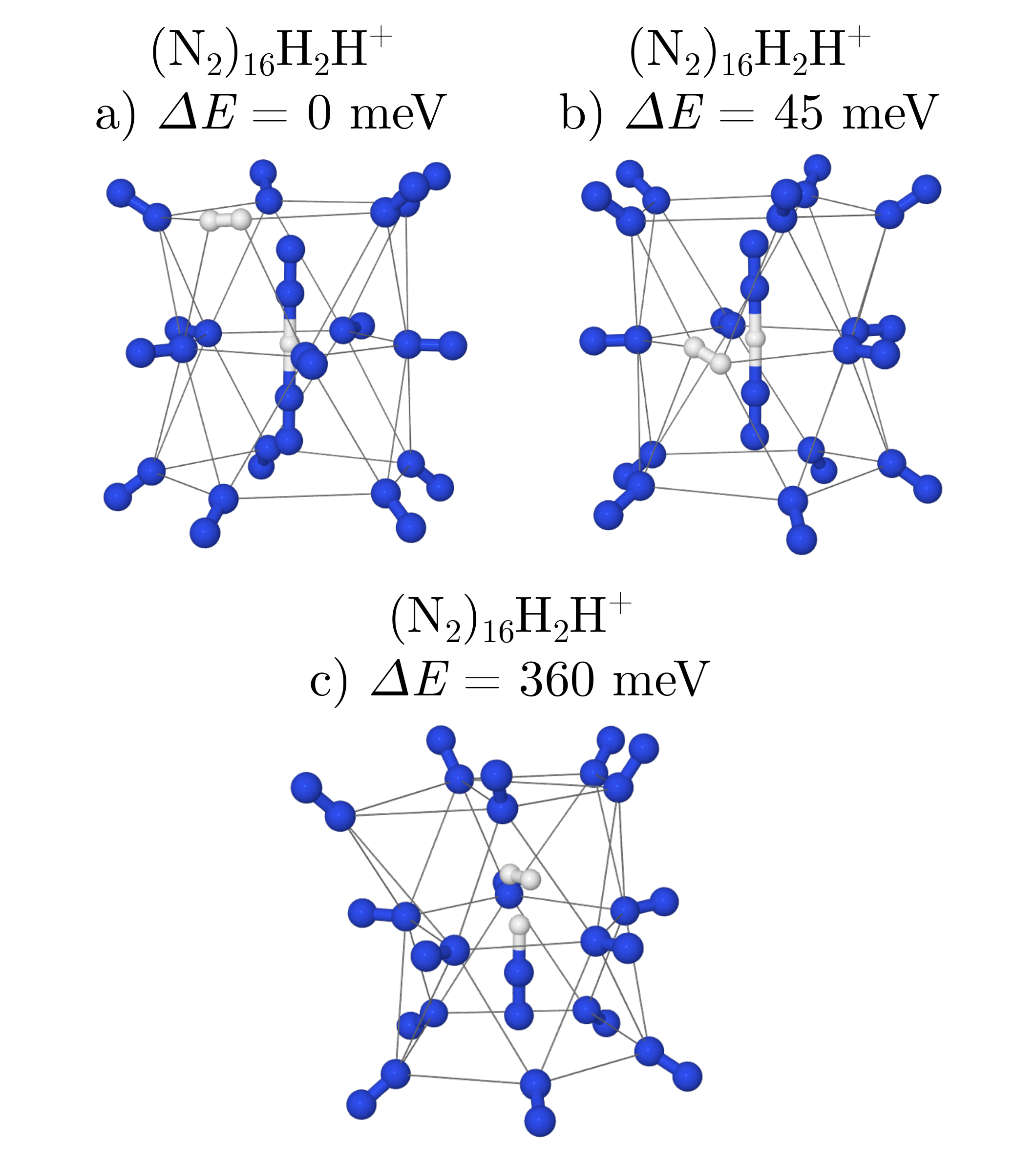}
\caption{Optimized structures  for the three main conformers of (N$_2$)$_{16}$H$_2$H$^+$ clusters at M06-HF/def2-TZVPP level of theory. The relative energies of the three systems are also shown. Structure a), with the neutral H$_2$ molecule in one of the outer pentagonal rings has the lowest potential energy, slightly lower than structure b) where this molecule occupies a position in the middle ring. Structure c) has the H$_2$ molecule as part of the central ionic core and exhibits both a much higher energy as well as a lopsided, less symmetric structure. The coordinates of the structures are given in the supplementary information.}
\label{fig:H2structs}
\end{figure}

\begin{figure}
\includegraphics[width=\columnwidth]{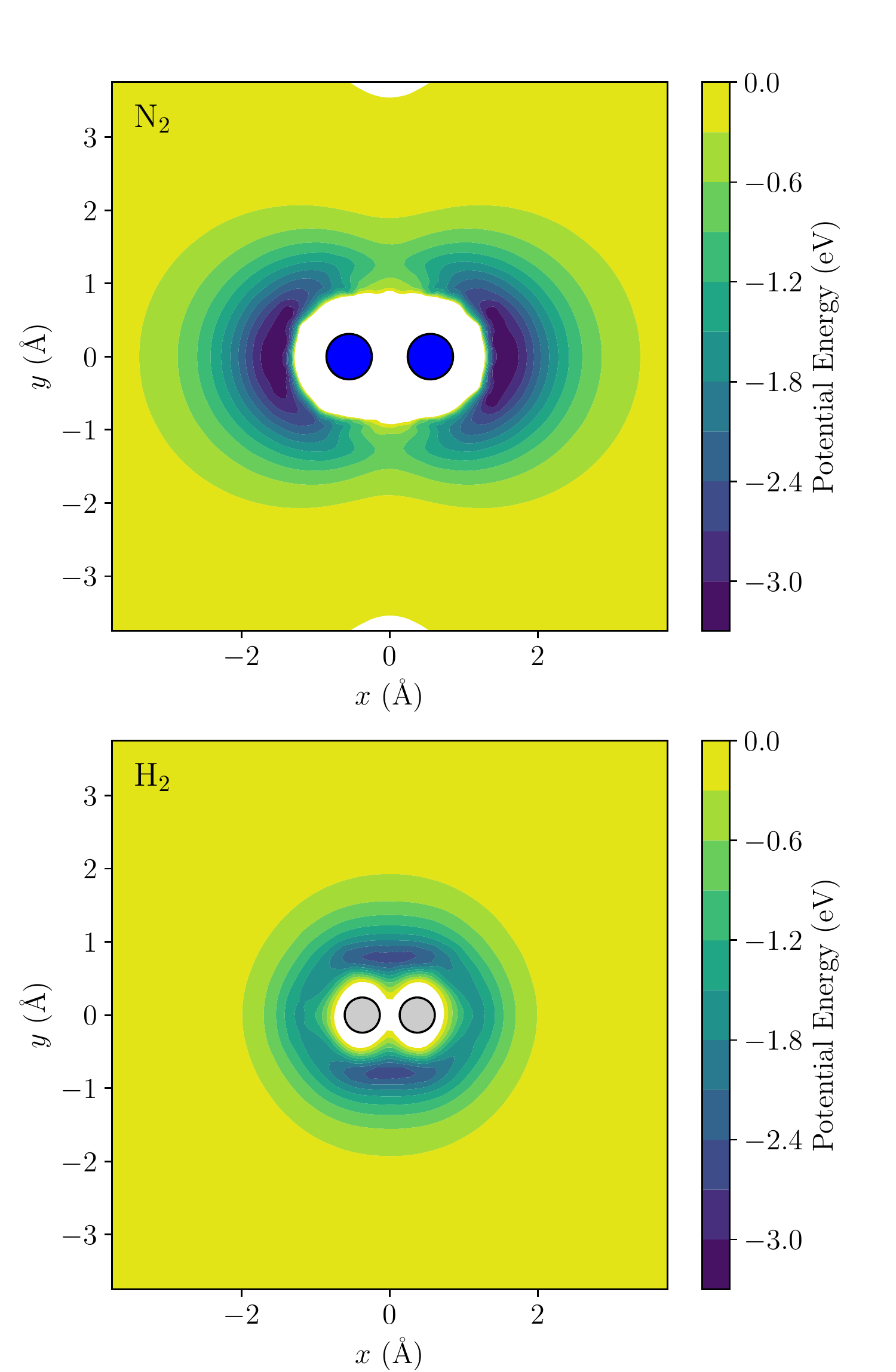}
\caption{Potential energy surfaces for a test point charge of 1\,e placed around N$_2$ and H$_2$, calculated at CCSD/cc-pVTZ level with frozen molecular structures (optimized without the point charge). Darker (bluer) colors indicate lower potential energy and positive energies are rendered as white.}
\label{fig:EPot}
\end{figure}

Calculated structures for three different conformers of (N$_2$)$_{16}$H$_2$H$^+$ (at M06-HF/def2-TZVPP level) are shown in Figure \ref{fig:H2structs}. All three are based on the structure of (N$_2$)$_{17}$H$^+$ shown in Figure \ref{fig:structs}, where a N$_2$ molecule has been replaced by H$_2$---due to symmetry these three replacements cover all possible combinations with the same basic structure---before the structures are re-optimized. In all cases it is clear that the overall structure of the cluster is largely preserved. However, unlike the N$_2$ molecules that are aligned with their molecular axes oriented towards the charge center, the H$_2$ molecule prefers to be aligned perpendicular to the charge. This can also be seen in Figure \ref{fig:EPot} where we show the interaction potential energy calculated for a point charge ($q=+1$\,e) around N$_2$ and H$_2$ molecules with frozen structures. In this picture, the interaction strength between the point charge and the molecules is similar in magnitude (minimum energy of 3.3\,eV for N$_2$ and 2.6\,eV for H$_2$), but it is clear that the preferred orientations are offset by 90 degrees. The completely different shapes of the potential energy surfaces for the two systems are the result of the orbitals involved in the interactions. For H$_2$, the point charge interacts with the only occupied molecular orbital ($\sigma_{1\text{s}}$), where the highest density is found between the H atoms. In the case of N$_2$, the charge interacts most favorably with the atomic 2s orbital of one of the atoms and near either end of the molecule.\cite{Papakondylis:2016aa} The interaction energies are consistent with the calculated proton affinities for the two molecules, at 5.1\,eV and 4.3\,eV, respectively at CCSD/cc-pVTZ level. They also match the known structures of the N$_2$H$^+$ and H$_3^+$ ions (linear and triangular, respectively).\cite{Papakondylis:2016aa,Carney:1976aa} The binding energy of an additional N$_2$ to the N$_2$H$^+$ ion is found to be 0.77\,eV, significantly higher than the binding energy of adding a H$_2$ molecule (0.25\,eV). This means that we should energetically expect the ionic core of the mixed clusters to consist of the (N$_2$-D-N$_2$)$^+$ ion, and the relative energies of the structures in Figure \ref{fig:H2structs} support this. The mixing of N$_2$ and H$_2$/D$_2$ would therefore predominantly take place amongst the loosely bound molecules surrounding the charged core.

Due to the complexity of the problem, we have refrained from calculating structures for higher order mixtures of nitrogen and hydrogen molecules than is shown in Figure \ref{fig:H2structs}. As the ratio of mixing increases, the number of possible conformers will drastically increase from the three that we have here. Nonetheless, some general observations arguments can be made about the mixing and the occurrence of magic numbers. The molecules surrounding the ionic core are attracted by ion-induced dipole interactions and as we have seen in Figures \ref{fig:H2structs} and \ref{fig:EPot}, the N$_2$ and H$_2$ prefer to align themselves perpendicular to each other relative to the charge center. The interactions between these neutral molecules will in turn be the result of non-bonding van der Waals forces. The typical interaction distances for such systems can be determined by the van der Waals radii of the atoms in the molecules. The van der Waals radii of N and H atoms are {1.55\,\AA} and 1.2\,\AA, respectively.\cite{Bondi:1964aa} The spacing between molecules when packing N$_2$, which are aligned with their axes oriented towards the center of the cluster, will then simply be given by twice the van der Waals radius, i.e.\ 3.1\,\AA. For the perpendicular H$_2$ on the other hand, the space occupied by a molecule in the lattice will be twice the van der Waals radius \emph{plus} the H-H bond length ($\sim 0.75$\,\AA), which gives 3.15\,\AA. Some studies have also found that the van der Waals radius of H is somewhat overestimated,\cite{Rowland:1996aa} which would reduce this number somewhat. Regardless, the interaction distances for the N$_2$ and perpendicular H$_2$ molecules are very similar and this is likely the reason for the efficient mixing of the two species since the overall structure is preserved when packing the molecules around the central ion. The reason why (N$_2$)$_n$(D$_2$)$_m$D$^+$ clusters where $n+m=17$ are  particularly stable with regards to mixing is likely due to this size representing the closure of the first solvation shell around the charge center. For sizes below this, the molecules will be more mobile within the cluster and there will less preference for a specific (and particularly stable) geometry for any given cluster size. For larger clusters the outer layers of molecules will be less influenced by the charge, reducing the preference of the N$_2$ and H/D$_2$ molecules to orient themselves perpendicular to each other. While these arguments explains the observations made, a more detailed explanation of the details will call for additional theory that lies beyond the scope of this work.



\section{Summary and Conclusions}

Like with rare gases,\cite{Gatchell:2018aa,Gatchell:2019ac} protonation has a strong effect on the overall structures and associated magic numbers of cationic N$_2$ clusters. The protonated clusters have a charge center consisting of a (N$_2$-H/D-N$_2$)$^+$ unit that extends the separation between these N$_2$ molecules compared to the N$_4^+$ unit in the pure clusters, and which affects the packing of neutral molecules around the center of the cluster. For the pure cationic clusters, abundance anomalies in the mass spectra indicate that the first solvation shell closes after a total of 19 N$_2$ molecules, while for the protonated systems this number is 17. 

Mixing of intact hydrogen molecules in the nitrogen clusters indicates that the two species are somewhat interchangeable with regards to the observed magic numbers. This is particularly true for protonated clusters consisting of a combined total of 17 N$_2$ and D$_2$ molecules, which are especially abundant for all measured mixtures. In our calculations we observe that the neutral N$_2$ molecules orient themselves with their axis pointing towards the charge center in the cluster, while the H/D$_2$ molecules are aligned perpendicular to this. In this orientation, the two different molecular species occupy the close to same amount of space in the first solvation shell, which explains the efficient mixing.

Due to the strong electrostrictive effect that the charge has on the first solvation layer, this shell can likely be considered to be solid. Comparable dense layers of localized solvation shells have been before, such as the ``snowballs'' of densely packed He atoms that can form around a range of ionic species.\cite{Atkins:1959aa,Buzzacchi:2001aa,Muller:2009ab,PhysRevLett.108.076101,Calvo:2015aa,Kuhn:2016aa,Rastogi:2018aa} The size-dependent interchangeability of the N$_2$ and H/D$_2$ molecules in solvation shells studied here show properties similar to substitutional alloys in metallurgy, as such they could be considered to be solid, dielectric nano-alloys.

\section*{Supplementary Material}
See supplementary material for coordinates of calculated structures shown in Figures \ref{fig:structs} and \ref{fig:H2structs}.

\begin{acknowledgments}
This work was supported by the Austrian Science Fund FWF (projects P31149 and P30355) and the Swedish Research Council (Contract No.\ 2016-06625).
\end{acknowledgments}


%
%

%


\bibliography{Library.bib}

\begin{thebibliography}{46}%
\makeatletter
\providecommand \@ifxundefined [1]{%
 \@ifx{#1\undefined}
}%
\providecommand \@ifnum [1]{%
 \ifnum #1\expandafter \@firstoftwo
 \else \expandafter \@secondoftwo
 \fi
}%
\providecommand \@ifx [1]{%
 \ifx #1\expandafter \@firstoftwo
 \else \expandafter \@secondoftwo
 \fi
}%
\providecommand \natexlab [1]{#1}%
\providecommand \enquote  [1]{``#1''}%
\providecommand \bibnamefont  [1]{#1}%
\providecommand \bibfnamefont [1]{#1}%
\providecommand \citenamefont [1]{#1}%
\providecommand \href@noop [0]{\@secondoftwo}%
\providecommand \href [0]{\begingroup \@sanitize@url \@href}%
\providecommand \@href[1]{\@@startlink{#1}\@@href}%
\providecommand \@@href[1]{\endgroup#1\@@endlink}%
\providecommand \@sanitize@url [0]{\catcode `\\12\catcode `\$12\catcode
  `\&12\catcode `\#12\catcode `\^12\catcode `\_12\catcode `\%12\relax}%
\providecommand \@@startlink[1]{}%
\providecommand \@@endlink[0]{}%
\providecommand \url  [0]{\begingroup\@sanitize@url \@url }%
\providecommand \@url [1]{\endgroup\@href {#1}{\urlprefix }}%
\providecommand \urlprefix  [0]{URL }%
\providecommand \Eprint [0]{\href }%
\providecommand \doibase [0]{http://dx.doi.org/}%
\providecommand \selectlanguage [0]{\@gobble}%
\providecommand \bibinfo  [0]{\@secondoftwo}%
\providecommand \bibfield  [0]{\@secondoftwo}%
\providecommand \translation [1]{[#1]}%
\providecommand \BibitemOpen [0]{}%
\providecommand \bibitemStop [0]{}%
\providecommand \bibitemNoStop [0]{.\EOS\space}%
\providecommand \EOS [0]{\spacefactor3000\relax}%
\providecommand \BibitemShut  [1]{\csname bibitem#1\endcsname}%
\let\auto@bib@innerbib\@empty
\bibitem [{\citenamefont {Ma}\ \emph {et~al.}(2017)\citenamefont {Ma},
  \citenamefont {Bao}, \citenamefont {Shi}, \citenamefont {Yan},\ and\
  \citenamefont {Zhang}}]{Ma:2017aa}%
  \BibitemOpen
  \bibfield  {author} {\bibinfo {author} {\bibfnamefont {J.-L.}\ \bibnamefont
  {Ma}}, \bibinfo {author} {\bibfnamefont {D.}~\bibnamefont {Bao}}, \bibinfo
  {author} {\bibfnamefont {M.-M.}\ \bibnamefont {Shi}}, \bibinfo {author}
  {\bibfnamefont {J.-M.}\ \bibnamefont {Yan}}, \ and\ \bibinfo {author}
  {\bibfnamefont {X.-B.}\ \bibnamefont {Zhang}},\ }\bibfield  {title} {\enquote
  {\bibinfo {title} {Reversible nitrogen fixation based on a rechargeable
  lithium-nitrogen battery for energy storage},}\ }\href {\doibase
  https://doi.org/10.1016/j.chempr.2017.03.016} {\bibfield  {journal} {\bibinfo
   {journal} {Chem}\ }\textbf {\bibinfo {volume} {2}},\ \bibinfo {pages}
  {525--532} (\bibinfo {year} {2017})}\BibitemShut {NoStop}%
\bibitem [{\citenamefont {Li}\ \emph {et~al.}(2018)\citenamefont {Li},
  \citenamefont {Feng}, \citenamefont {Liu}, \citenamefont {Hao}, \citenamefont
  {Redfern}, \citenamefont {Lei}, \citenamefont {Liu},\ and\ \citenamefont
  {Ma}}]{Li:2018aa}%
  \BibitemOpen
  \bibfield  {author} {\bibinfo {author} {\bibfnamefont {Y.}~\bibnamefont
  {Li}}, \bibinfo {author} {\bibfnamefont {X.}~\bibnamefont {Feng}}, \bibinfo
  {author} {\bibfnamefont {H.}~\bibnamefont {Liu}}, \bibinfo {author}
  {\bibfnamefont {J.}~\bibnamefont {Hao}}, \bibinfo {author} {\bibfnamefont
  {S.~A.~T.}\ \bibnamefont {Redfern}}, \bibinfo {author} {\bibfnamefont
  {W.}~\bibnamefont {Lei}}, \bibinfo {author} {\bibfnamefont {D.}~\bibnamefont
  {Liu}}, \ and\ \bibinfo {author} {\bibfnamefont {Y.}~\bibnamefont {Ma}},\
  }\bibfield  {title} {\enquote {\bibinfo {title} {Route to high-energy density
  polymeric nitrogen t-n via he−n compounds},}\ }\href {\doibase
  10.1038/s41467-018-03200-4} {\bibfield  {journal} {\bibinfo  {journal}
  {Nature Communications}\ }\textbf {\bibinfo {volume} {9}},\ \bibinfo {pages}
  {722} (\bibinfo {year} {2018})}\BibitemShut {NoStop}%
\bibitem [{\citenamefont {Xu}\ and\ \citenamefont {Dunning}(2016)}]{Xu:2016aa}%
  \BibitemOpen
  \bibfield  {author} {\bibinfo {author} {\bibfnamefont {L.~T.}\ \bibnamefont
  {Xu}}\ and\ \bibinfo {author} {\bibfnamefont {T.~H.}\ \bibnamefont
  {Dunning}},\ }\bibfield  {title} {\enquote {\bibinfo {title} {Variations in
  the nature of triple bonds: The n2, hcn, and hc2h series},}\ }\bibfield
  {booktitle} {\emph {\bibinfo {booktitle} {The Journal of Physical Chemistry
  A}},\ }\href {\doibase 10.1021/acs.jpca.6b03631} {\bibfield  {journal}
  {\bibinfo  {journal} {The Journal of Physical Chemistry A}\ }\textbf
  {\bibinfo {volume} {120}},\ \bibinfo {pages} {4526--4533} (\bibinfo {year}
  {2016})}\BibitemShut {NoStop}%
\bibitem [{\citenamefont {Samartzis}\ and\ \citenamefont
  {Wodtke}(2006)}]{Samartzis:2006aa}%
  \BibitemOpen
  \bibfield  {author} {\bibinfo {author} {\bibfnamefont {P.~C.}\ \bibnamefont
  {Samartzis}}\ and\ \bibinfo {author} {\bibfnamefont {A.~M.}\ \bibnamefont
  {Wodtke}},\ }\bibfield  {title} {\enquote {\bibinfo {title} {All-nitrogen
  chemistry: how far are we from n60?}}\ }\bibfield  {booktitle} {\emph
  {\bibinfo {booktitle} {International Reviews in Physical Chemistry}},\ }\href
  {\doibase 10.1080/01442350600879319} {\bibfield  {journal} {\bibinfo
  {journal} {International Reviews in Physical Chemistry}\ }\textbf {\bibinfo
  {volume} {25}},\ \bibinfo {pages} {527--552} (\bibinfo {year}
  {2006})}\BibitemShut {NoStop}%
\bibitem [{\citenamefont {Jonkman}\ and\ \citenamefont
  {Michl}(1981)}]{Jonkman:1981aa}%
  \BibitemOpen
  \bibfield  {author} {\bibinfo {author} {\bibfnamefont {H.~T.}\ \bibnamefont
  {Jonkman}}\ and\ \bibinfo {author} {\bibfnamefont {J.}~\bibnamefont
  {Michl}},\ }\bibfield  {title} {\enquote {\bibinfo {title} {Secondary ion
  mass spectrometry of small-molecule solids at cryogenic temperatures. 1.
  nitrogen and carbon monoxide},}\ }\bibfield  {booktitle} {\emph {\bibinfo
  {booktitle} {Journal of the American Chemical Society}},\ }\href {\doibase
  10.1021/ja00394a001} {\bibfield  {journal} {\bibinfo  {journal} {Journal of
  the American Chemical Society}\ }\textbf {\bibinfo {volume} {103}},\ \bibinfo
  {pages} {733--737} (\bibinfo {year} {1981})}\BibitemShut {NoStop}%
\bibitem [{\citenamefont {Tonuma}\ \emph {et~al.}(1994)\citenamefont {Tonuma},
  \citenamefont {Kumagai}, \citenamefont {Matsuo}, \citenamefont {Shibata},\
  and\ \citenamefont {Tawara}}]{Tonuma:1994aa}%
  \BibitemOpen
  \bibfield  {author} {\bibinfo {author} {\bibfnamefont {T.}~\bibnamefont
  {Tonuma}}, \bibinfo {author} {\bibfnamefont {H.}~\bibnamefont {Kumagai}},
  \bibinfo {author} {\bibfnamefont {T.}~\bibnamefont {Matsuo}}, \bibinfo
  {author} {\bibfnamefont {H.}~\bibnamefont {Shibata}}, \ and\ \bibinfo
  {author} {\bibfnamefont {H.}~\bibnamefont {Tawara}},\ }\bibfield  {title}
  {\enquote {\bibinfo {title} {Positive and negative cluster ions and multiply
  charged ions produced from frozen nitrogen, carbon monoxide and oxygen
  molecules under energetic, heavy-ion impact},}\ }\href {\doibase
  https://doi.org/10.1016/0168-1176(94)04002-8} {\bibfield  {journal} {\bibinfo
   {journal} {International Journal of Mass Spectrometry and Ion Processes}\
  }\textbf {\bibinfo {volume} {135}},\ \bibinfo {pages} {129--137} (\bibinfo
  {year} {1994})}\BibitemShut {NoStop}%
\bibitem [{\citenamefont {Fern{\'a}ndez-Lima}\ \emph
  {et~al.}(2007)\citenamefont {Fern{\'a}ndez-Lima}, \citenamefont {Ponciano},
  \citenamefont {Faraudo}, \citenamefont {Grivet}, \citenamefont
  {da~Silveira},\ and\ \citenamefont {Nascimento}}]{Fernandez-Lima:2007aa}%
  \BibitemOpen
  \bibfield  {author} {\bibinfo {author} {\bibfnamefont {F.~A.}\ \bibnamefont
  {Fern{\'a}ndez-Lima}}, \bibinfo {author} {\bibfnamefont {C.~R.}\ \bibnamefont
  {Ponciano}}, \bibinfo {author} {\bibfnamefont {G.~S.}\ \bibnamefont
  {Faraudo}}, \bibinfo {author} {\bibfnamefont {M.}~\bibnamefont {Grivet}},
  \bibinfo {author} {\bibfnamefont {E.~F.}\ \bibnamefont {da~Silveira}}, \ and\
  \bibinfo {author} {\bibfnamefont {M.~A.~C.}\ \bibnamefont {Nascimento}},\
  }\bibfield  {title} {\enquote {\bibinfo {title} {Characterization of
  nn=2--18+ clusters produced by 252cf fission fragment impact on a n2 ice
  target},}\ }\href {\doibase https://doi.org/10.1016/j.chemphys.2007.08.022}
  {\bibfield  {journal} {\bibinfo  {journal} {Chemical Physics}\ }\textbf
  {\bibinfo {volume} {340}},\ \bibinfo {pages} {127--133} (\bibinfo {year}
  {2007})}\BibitemShut {NoStop}%
\bibitem [{\citenamefont {Ponciano}\ \emph {et~al.}(2008)\citenamefont
  {Ponciano}, \citenamefont {Martinez}, \citenamefont {Farenzena},
  \citenamefont {Iza}, \citenamefont {Homem}, \citenamefont {Naves~de Brito},
  \citenamefont {Wien},\ and\ \citenamefont {da~Silveira}}]{Ponciano:2008aa}%
  \BibitemOpen
  \bibfield  {author} {\bibinfo {author} {\bibfnamefont {C.~R.}\ \bibnamefont
  {Ponciano}}, \bibinfo {author} {\bibfnamefont {R.}~\bibnamefont {Martinez}},
  \bibinfo {author} {\bibfnamefont {L.~S.}\ \bibnamefont {Farenzena}}, \bibinfo
  {author} {\bibfnamefont {P.}~\bibnamefont {Iza}}, \bibinfo {author}
  {\bibfnamefont {M.~G.~P.}\ \bibnamefont {Homem}}, \bibinfo {author}
  {\bibfnamefont {A.}~\bibnamefont {Naves~de Brito}}, \bibinfo {author}
  {\bibfnamefont {K.}~\bibnamefont {Wien}}, \ and\ \bibinfo {author}
  {\bibfnamefont {E.~F.}\ \bibnamefont {da~Silveira}},\ }\bibfield  {title}
  {\enquote {\bibinfo {title} {Cluster emission and chemical reactions in
  oxygen and nitrogen ices induced by fast heavy-ion impact},}\ }\bibfield
  {booktitle} {\emph {\bibinfo {booktitle} {Journal of Mass Spectrometry}},\
  }\href {\doibase 10.1002/jms.1429} {\bibfield  {journal} {\bibinfo  {journal}
  {Journal of Mass Spectrometry}\ }\textbf {\bibinfo {volume} {43}},\ \bibinfo
  {pages} {1521--1530} (\bibinfo {year} {2008})}\BibitemShut {NoStop}%
\bibitem [{\citenamefont {Friedman}\ and\ \citenamefont
  {Beuhler}(1983)}]{Friedman:1983aa}%
  \BibitemOpen
  \bibfield  {author} {\bibinfo {author} {\bibfnamefont {L.}~\bibnamefont
  {Friedman}}\ and\ \bibinfo {author} {\bibfnamefont {R.~J.}\ \bibnamefont
  {Beuhler}},\ }\bibfield  {title} {\enquote {\bibinfo {title} {Magic numbers
  for argon and nitrogen cluster ions},}\ }\bibfield  {booktitle} {\emph
  {\bibinfo {booktitle} {The Journal of Chemical Physics}},\ }\href {\doibase
  10.1063/1.445312} {\bibfield  {journal} {\bibinfo  {journal} {The Journal of
  Chemical Physics}\ }\textbf {\bibinfo {volume} {78}},\ \bibinfo {pages}
  {4669--4675} (\bibinfo {year} {1983})}\BibitemShut {NoStop}%
\bibitem [{\citenamefont {Scheier}\ and\ \citenamefont
  {M{\"a}rk}(1988)}]{Scheier:1988ab}%
  \BibitemOpen
  \bibfield  {author} {\bibinfo {author} {\bibfnamefont {P.}~\bibnamefont
  {Scheier}}\ and\ \bibinfo {author} {\bibfnamefont {T.~D.}\ \bibnamefont
  {M{\"a}rk}},\ }\bibfield  {title} {\enquote {\bibinfo {title} {Quantized
  sequential decay series of metastable n2 cluster ions (n2)n+* →(n2)n-1+*
  →{\ldots}→n2+},}\ }\href {\doibase
  https://doi.org/10.1016/0009-2614(88)87193-8} {\bibfield  {journal} {\bibinfo
   {journal} {Chemical Physics Letters}\ }\textbf {\bibinfo {volume} {148}},\
  \bibinfo {pages} {393--400} (\bibinfo {year} {1988})}\BibitemShut {NoStop}%
\bibitem [{\citenamefont {Leisner}\ \emph {et~al.}(1988)\citenamefont
  {Leisner}, \citenamefont {Echt}, \citenamefont {Kandler}, \citenamefont
  {Yan},\ and\ \citenamefont {Recknagel}}]{Leisner:1988aa}%
  \BibitemOpen
  \bibfield  {author} {\bibinfo {author} {\bibfnamefont {T.}~\bibnamefont
  {Leisner}}, \bibinfo {author} {\bibfnamefont {O.}~\bibnamefont {Echt}},
  \bibinfo {author} {\bibfnamefont {O.}~\bibnamefont {Kandler}}, \bibinfo
  {author} {\bibfnamefont {X.-J.}\ \bibnamefont {Yan}}, \ and\ \bibinfo
  {author} {\bibfnamefont {E.}~\bibnamefont {Recknagel}},\ }\bibfield  {title}
  {\enquote {\bibinfo {title} {Quantum effects in the decomposition of nitrogen
  cluster ions},}\ }\href {\doibase
  https://doi.org/10.1016/0009-2614(88)87192-6} {\bibfield  {journal} {\bibinfo
   {journal} {Chemical Physics Letters}\ }\textbf {\bibinfo {volume} {148}},\
  \bibinfo {pages} {386--392} (\bibinfo {year} {1988})}\BibitemShut {NoStop}%
\bibitem [{\citenamefont {Scheier}, \citenamefont {Stamatovic},\ and\
  \citenamefont {M{\"a}rk}(1988)}]{Scheier:1988aa}%
  \BibitemOpen
  \bibfield  {author} {\bibinfo {author} {\bibfnamefont {P.}~\bibnamefont
  {Scheier}}, \bibinfo {author} {\bibfnamefont {A.}~\bibnamefont {Stamatovic}},
  \ and\ \bibinfo {author} {\bibfnamefont {T.~D.}\ \bibnamefont {M{\"a}rk}},\
  }\bibfield  {title} {\enquote {\bibinfo {title} {Production and properties of
  singly, doubly, and triply charged n2 clusters},}\ }\bibfield  {booktitle}
  {\emph {\bibinfo {booktitle} {The Journal of Chemical Physics}},\ }\href
  {\doibase 10.1063/1.453787} {\bibfield  {journal} {\bibinfo  {journal} {The
  Journal of Chemical Physics}\ }\textbf {\bibinfo {volume} {88}},\ \bibinfo
  {pages} {4289--4293} (\bibinfo {year} {1988})}\BibitemShut {NoStop}%
\bibitem [{\citenamefont {M{\"a}rk}\ \emph {et~al.}(1989)\citenamefont
  {M{\"a}rk}, \citenamefont {Scheier}, \citenamefont {Lezius}, \citenamefont
  {Walder},\ and\ \citenamefont {Stamatovic}}]{Mark:1989aa}%
  \BibitemOpen
  \bibfield  {author} {\bibinfo {author} {\bibfnamefont {T.~D.}\ \bibnamefont
  {M{\"a}rk}}, \bibinfo {author} {\bibfnamefont {P.}~\bibnamefont {Scheier}},
  \bibinfo {author} {\bibfnamefont {M.}~\bibnamefont {Lezius}}, \bibinfo
  {author} {\bibfnamefont {G.}~\bibnamefont {Walder}}, \ and\ \bibinfo {author}
  {\bibfnamefont {A.}~\bibnamefont {Stamatovic}},\ }\bibfield  {title}
  {\enquote {\bibinfo {title} {Multiply charged cluster ions of ar, kr, xe, n2,
  o2, co2 so2 and nh3: Production mechanism, appearance size and appearance
  energy},}\ }in\ \href@noop {} {\emph {\bibinfo {booktitle} {Small Particles
  and Inorganic Clusters}}},\ \bibinfo {editor} {edited by\ \bibinfo {editor}
  {\bibfnamefont {C.}~\bibnamefont {Chapon}}, \bibinfo {editor} {\bibfnamefont
  {M.~F.}\ \bibnamefont {Gillet}}, \ and\ \bibinfo {editor} {\bibfnamefont
  {C.~R.}\ \bibnamefont {Henry}}}\ (\bibinfo  {publisher} {Springer Berlin
  Heidelberg},\ \bibinfo {address} {Berlin, Heidelberg},\ \bibinfo {year}
  {1989})\ pp.\ \bibinfo {pages} {279--281}\BibitemShut {NoStop}%
\bibitem [{\citenamefont {Walder}\ \emph {et~al.}(1991)\citenamefont {Walder},
  \citenamefont {Foltin}, \citenamefont {Stefanson}, \citenamefont
  {Castleman},\ and\ \citenamefont {M{\"a}rk}}]{Walder:1991aa}%
  \BibitemOpen
  \bibfield  {author} {\bibinfo {author} {\bibfnamefont {G.}~\bibnamefont
  {Walder}}, \bibinfo {author} {\bibfnamefont {M.}~\bibnamefont {Foltin}},
  \bibinfo {author} {\bibfnamefont {T.}~\bibnamefont {Stefanson}}, \bibinfo
  {author} {\bibfnamefont {A.~W.}\ \bibnamefont {Castleman}}, \ and\ \bibinfo
  {author} {\bibfnamefont {T.~D.}\ \bibnamefont {M{\"a}rk}},\ }\bibfield
  {title} {\enquote {\bibinfo {title} {Metastable decay of stoichiometric and
  non-stoichiometric nitrogen cluster ions},}\ }\href {\doibase
  https://doi.org/10.1016/0168-1176(91)85077-Y} {\bibfield  {journal} {\bibinfo
   {journal} {International Journal of Mass Spectrometry and Ion Processes}\
  }\textbf {\bibinfo {volume} {107}},\ \bibinfo {pages} {127--134} (\bibinfo
  {year} {1991})}\BibitemShut {NoStop}%
\bibitem [{\citenamefont {Flesch}\ \emph {et~al.}(2004)\citenamefont {Flesch},
  \citenamefont {Kosugi}, \citenamefont {Bradeanu}, \citenamefont {Neville},\
  and\ \citenamefont {R{\"u}hl}}]{Flesch:2004aa}%
  \BibitemOpen
  \bibfield  {author} {\bibinfo {author} {\bibfnamefont {R.}~\bibnamefont
  {Flesch}}, \bibinfo {author} {\bibfnamefont {N.}~\bibnamefont {Kosugi}},
  \bibinfo {author} {\bibfnamefont {I.~L.}\ \bibnamefont {Bradeanu}}, \bibinfo
  {author} {\bibfnamefont {J.~J.}\ \bibnamefont {Neville}}, \ and\ \bibinfo
  {author} {\bibfnamefont {E.}~\bibnamefont {R{\"u}hl}},\ }\bibfield  {title}
  {\enquote {\bibinfo {title} {Cluster size effects in core excitons of
  1s-excited nitrogen},}\ }\bibfield  {booktitle} {\emph {\bibinfo {booktitle}
  {The Journal of Chemical Physics}},\ }\href {\doibase 10.1063/1.1804180}
  {\bibfield  {journal} {\bibinfo  {journal} {The Journal of Chemical Physics}\
  }\textbf {\bibinfo {volume} {121}},\ \bibinfo {pages} {8343--8350} (\bibinfo
  {year} {2004})}\BibitemShut {NoStop}%
\bibitem [{\citenamefont {Nguyen}\ and\ \citenamefont
  {Ha}(2001)}]{Nguyen:2001aa}%
  \BibitemOpen
  \bibfield  {author} {\bibinfo {author} {\bibfnamefont {M.~T.}\ \bibnamefont
  {Nguyen}}\ and\ \bibinfo {author} {\bibfnamefont {T.-K.}\ \bibnamefont
  {Ha}},\ }\bibfield  {title} {\enquote {\bibinfo {title} {Decomposition
  mechanism of the polynitrogen n5 and n6 clusters and their ions},}\ }\href
  {\doibase https://doi.org/10.1016/S0009-2614(01)00037-9} {\bibfield
  {journal} {\bibinfo  {journal} {Chemical Physics Letters}\ }\textbf {\bibinfo
  {volume} {335}},\ \bibinfo {pages} {311--320} (\bibinfo {year}
  {2001})}\BibitemShut {NoStop}%
\bibitem [{\citenamefont {Li}\ and\ \citenamefont {Zhao}(2002)}]{Li:2002aa}%
  \BibitemOpen
  \bibfield  {author} {\bibinfo {author} {\bibfnamefont {Q.~S.}\ \bibnamefont
  {Li}}\ and\ \bibinfo {author} {\bibfnamefont {J.~F.}\ \bibnamefont {Zhao}},\
  }\bibfield  {title} {\enquote {\bibinfo {title} {A theoretical study on
  decomposition pathways of n7+ and n7- clusters},}\ }\bibfield  {booktitle}
  {\emph {\bibinfo {booktitle} {The Journal of Physical Chemistry A}},\ }\href
  {\doibase 10.1021/jp014402k} {\bibfield  {journal} {\bibinfo  {journal} {The
  Journal of Physical Chemistry A}\ }\textbf {\bibinfo {volume} {106}},\
  \bibinfo {pages} {5928--5931} (\bibinfo {year} {2002})}\BibitemShut {NoStop}%
\bibitem [{\citenamefont {Evangelisti}\ and\ \citenamefont
  {Leininger}(2003)}]{Evangelisti:2003aa}%
  \BibitemOpen
  \bibfield  {author} {\bibinfo {author} {\bibfnamefont {S.}~\bibnamefont
  {Evangelisti}}\ and\ \bibinfo {author} {\bibfnamefont {T.}~\bibnamefont
  {Leininger}},\ }\bibfield  {title} {\enquote {\bibinfo {title} {Ionic
  nitrogen clusters},}\ }\bibfield  {booktitle} {\emph {\bibinfo {booktitle}
  {2001 Quitel S.I.}},\ }\href {\doibase
  https://doi.org/10.1016/S0166-1280(02)00532-8} {\bibfield  {journal}
  {\bibinfo  {journal} {Journal of Molecular Structure: THEOCHEM}\ }\textbf
  {\bibinfo {volume} {621}},\ \bibinfo {pages} {43--50} (\bibinfo {year}
  {2003})}\BibitemShut {NoStop}%
\bibitem [{\citenamefont {Law}\ \emph {et~al.}(2002)\citenamefont {Law},
  \citenamefont {Li}, \citenamefont {Wang}, \citenamefont {Tian},\ and\
  \citenamefont {Wong}}]{Law:2002aa}%
  \BibitemOpen
  \bibfield  {author} {\bibinfo {author} {\bibfnamefont {C.~K.}\ \bibnamefont
  {Law}}, \bibinfo {author} {\bibfnamefont {W.~K.}\ \bibnamefont {Li}},
  \bibinfo {author} {\bibfnamefont {X.}~\bibnamefont {Wang}}, \bibinfo {author}
  {\bibfnamefont {A.}~\bibnamefont {Tian}}, \ and\ \bibinfo {author}
  {\bibfnamefont {N.~B.}\ \bibnamefont {Wong}},\ }\bibfield  {title} {\enquote
  {\bibinfo {title} {A gaussian-3 study of n7+ and n7−isomers},}\ }\href
  {\doibase https://doi.org/10.1016/S0166-1280(02)00411-6} {\bibfield
  {journal} {\bibinfo  {journal} {Journal of Molecular Structure: THEOCHEM}\
  }\textbf {\bibinfo {volume} {617}},\ \bibinfo {pages} {121--131} (\bibinfo
  {year} {2002})}\BibitemShut {NoStop}%
\bibitem [{\citenamefont {Nowak}, \citenamefont {Menapace},\ and\ \citenamefont
  {Bernstein}(1988)}]{Nowak:1988aa}%
  \BibitemOpen
  \bibfield  {author} {\bibinfo {author} {\bibfnamefont {R.}~\bibnamefont
  {Nowak}}, \bibinfo {author} {\bibfnamefont {J.~A.}\ \bibnamefont {Menapace}},
  \ and\ \bibinfo {author} {\bibfnamefont {E.~R.}\ \bibnamefont {Bernstein}},\
  }\bibfield  {title} {\enquote {\bibinfo {title} {Benzene clustered with n2,
  co2, and co: Energy levels, vibrational structure, and nucleation},}\
  }\bibfield  {booktitle} {\emph {\bibinfo {booktitle} {The Journal of Chemical
  Physics}},\ }\href {\doibase 10.1063/1.455182} {\bibfield  {journal}
  {\bibinfo  {journal} {The Journal of Chemical Physics}\ }\textbf {\bibinfo
  {volume} {89}},\ \bibinfo {pages} {1309--1321} (\bibinfo {year}
  {1988})}\BibitemShut {NoStop}%
\bibitem [{\citenamefont {Baer}, \citenamefont {Marx},\ and\ \citenamefont
  {Mathias}(2010)}]{Baer:2010aa}%
  \BibitemOpen
  \bibfield  {author} {\bibinfo {author} {\bibfnamefont {M.}~\bibnamefont
  {Baer}}, \bibinfo {author} {\bibfnamefont {D.}~\bibnamefont {Marx}}, \ and\
  \bibinfo {author} {\bibfnamefont {G.}~\bibnamefont {Mathias}},\ }\bibfield
  {title} {\enquote {\bibinfo {title} {Theoretical messenger spectroscopy of
  microsolvated hydronium and zundel cations},}\ }\bibfield  {booktitle} {\emph
  {\bibinfo {booktitle} {Angewandte Chemie International Edition}},\ }\href
  {\doibase 10.1002/anie.201001672} {\bibfield  {journal} {\bibinfo  {journal}
  {Angewandte Chemie International Edition}\ }\textbf {\bibinfo {volume}
  {49}},\ \bibinfo {pages} {7346--7349} (\bibinfo {year} {2010})}\BibitemShut
  {NoStop}%
\bibitem [{\citenamefont {Craig}, \citenamefont {Menges},\ and\ \citenamefont
  {Johnson}(2017)}]{Craig:2017aa}%
  \BibitemOpen
  \bibfield  {author} {\bibinfo {author} {\bibfnamefont {S.~M.}\ \bibnamefont
  {Craig}}, \bibinfo {author} {\bibfnamefont {F.~S.}\ \bibnamefont {Menges}}, \
  and\ \bibinfo {author} {\bibfnamefont {M.~A.}\ \bibnamefont {Johnson}},\
  }\bibfield  {title} {\enquote {\bibinfo {title} {Application of gas phase
  cryogenic vibrational spectroscopy to characterize the co2, co, n2 and n2o
  interactions with the open coordination site on a ni(i) macrocycle using dual
  cryogenic ion traps},}\ }\bibfield  {booktitle} {\emph {\bibinfo {booktitle}
  {Molecular Spectroscopy in Traps}},\ }\href {\doibase
  https://doi.org/10.1016/j.jms.2016.11.015} {\bibfield  {journal} {\bibinfo
  {journal} {Journal of Molecular Spectroscopy}\ }\textbf {\bibinfo {volume}
  {332}},\ \bibinfo {pages} {117--123} (\bibinfo {year} {2017})}\BibitemShut
  {NoStop}%
\bibitem [{\citenamefont {Weinberger}\ \emph {et~al.}(2017)\citenamefont
  {Weinberger}, \citenamefont {Postler}, \citenamefont {Scheier},\ and\
  \citenamefont {Echt}}]{Weinberger:2017aa}%
  \BibitemOpen
  \bibfield  {author} {\bibinfo {author} {\bibfnamefont {N.}~\bibnamefont
  {Weinberger}}, \bibinfo {author} {\bibfnamefont {J.}~\bibnamefont {Postler}},
  \bibinfo {author} {\bibfnamefont {P.}~\bibnamefont {Scheier}}, \ and\
  \bibinfo {author} {\bibfnamefont {O.}~\bibnamefont {Echt}},\ }\bibfield
  {title} {\enquote {\bibinfo {title} {Nitrogen cluster anions},}\ }\bibfield
  {booktitle} {\emph {\bibinfo {booktitle} {The Journal of Physical Chemistry
  C}},\ }\href {\doibase 10.1021/acs.jpcc.6b09211} {\bibfield  {journal}
  {\bibinfo  {journal} {The Journal of Physical Chemistry C}\ }\textbf
  {\bibinfo {volume} {121}},\ \bibinfo {pages} {10632--10637} (\bibinfo {year}
  {2017})}\BibitemShut {NoStop}%
\bibitem [{\citenamefont {Vostrikov}\ and\ \citenamefont
  {Dubov}(2006)}]{Vostrikov:2006aa}%
  \BibitemOpen
  \bibfield  {author} {\bibinfo {author} {\bibfnamefont {A.~A.}\ \bibnamefont
  {Vostrikov}}\ and\ \bibinfo {author} {\bibfnamefont {D.~Y.}\ \bibnamefont
  {Dubov}},\ }\bibfield  {title} {\enquote {\bibinfo {title} {Absolute cross
  sections of electron attachment to molecular clusters. part ii: Formation of
  (h2o)n−, (n2o)n−, and (n2)n−},}\ }\href {\doibase
  10.1134/S1063784206120012} {\bibfield  {journal} {\bibinfo  {journal}
  {Technical Physics}\ }\textbf {\bibinfo {volume} {51}},\ \bibinfo {pages}
  {1537--1552} (\bibinfo {year} {2006})}\BibitemShut {NoStop}%
\bibitem [{\citenamefont {Yurtsever}\ and\ \citenamefont
  {Calvo}(2019)}]{Yurtsever:2019aa}%
  \BibitemOpen
  \bibfield  {author} {\bibinfo {author} {\bibfnamefont {E.}~\bibnamefont
  {Yurtsever}}\ and\ \bibinfo {author} {\bibfnamefont {F.}~\bibnamefont
  {Calvo}},\ }\bibfield  {title} {\enquote {\bibinfo {title} {Quantum chemical
  view on the growth mechanisms of odd-sized nitrogen cluster anions},}\
  }\bibfield  {booktitle} {\emph {\bibinfo {booktitle} {The Journal of Physical
  Chemistry A}},\ }\href {\doibase 10.1021/acs.jpca.8b08822} {\bibfield
  {journal} {\bibinfo  {journal} {The Journal of Physical Chemistry A}\
  }\textbf {\bibinfo {volume} {123}},\ \bibinfo {pages} {202--209} (\bibinfo
  {year} {2019})}\BibitemShut {NoStop}%
\bibitem [{\citenamefont {Calvo}, \citenamefont {Boutin},\ and\ \citenamefont
  {Labastie}(1999)}]{Calvo:1999aa}%
  \BibitemOpen
  \bibfield  {author} {\bibinfo {author} {\bibfnamefont {F.}~\bibnamefont
  {Calvo}}, \bibinfo {author} {\bibfnamefont {A.}~\bibnamefont {Boutin}}, \
  and\ \bibinfo {author} {\bibfnamefont {P.}~\bibnamefont {Labastie}},\
  }\bibfield  {title} {\enquote {\bibinfo {title} {Structure of nitrogen
  molecular clusters (n2)n with 13≤n≤55},}\ }\href {\doibase
  10.1007/s100530050424} {\bibfield  {journal} {\bibinfo  {journal} {The
  European Physical Journal D - Atomic, Molecular, Optical and Plasma Physics}\
  }\textbf {\bibinfo {volume} {9}},\ \bibinfo {pages} {189--193} (\bibinfo
  {year} {1999})}\BibitemShut {NoStop}%
\bibitem [{\citenamefont {Echt}, \citenamefont {Sattler},\ and\ \citenamefont
  {Recknagel}(1981)}]{Echt:1981aa}%
  \BibitemOpen
  \bibfield  {author} {\bibinfo {author} {\bibfnamefont {O.}~\bibnamefont
  {Echt}}, \bibinfo {author} {\bibfnamefont {K.}~\bibnamefont {Sattler}}, \
  and\ \bibinfo {author} {\bibfnamefont {E.}~\bibnamefont {Recknagel}},\
  }\bibfield  {title} {\enquote {\bibinfo {title} {Magic numbers for sphere
  packings: Experimental verification in free xenon clusters},}\ }\href@noop {}
  {\bibfield  {journal} {\bibinfo  {journal} {Physical Review Letters}\
  }\textbf {\bibinfo {volume} {47}},\ \bibinfo {pages} {1121--1124} (\bibinfo
  {year} {1981})}\BibitemShut {NoStop}%
\bibitem [{\citenamefont {Harris}\ \emph {et~al.}(1986)\citenamefont {Harris},
  \citenamefont {Norman}, \citenamefont {Mulkern},\ and\ \citenamefont
  {Northby}}]{Harris:1986aa}%
  \BibitemOpen
  \bibfield  {author} {\bibinfo {author} {\bibfnamefont {I.~A.}\ \bibnamefont
  {Harris}}, \bibinfo {author} {\bibfnamefont {K.~A.}\ \bibnamefont {Norman}},
  \bibinfo {author} {\bibfnamefont {R.~V.}\ \bibnamefont {Mulkern}}, \ and\
  \bibinfo {author} {\bibfnamefont {J.~A.}\ \bibnamefont {Northby}},\
  }\bibfield  {title} {\enquote {\bibinfo {title} {Icosahedral structure of
  large charged argon clusters},}\ }\href@noop {} {\bibfield  {journal}
  {\bibinfo  {journal} {Chemical Physics Letters}\ }\textbf {\bibinfo {volume}
  {130}},\ \bibinfo {pages} {316--320} (\bibinfo {year} {1986})}\BibitemShut
  {NoStop}%
\bibitem [{\citenamefont {Scheier}\ and\ \citenamefont
  {M{\"a}rk}(1987)}]{Scheier:1987aa}%
  \BibitemOpen
  \bibfield  {author} {\bibinfo {author} {\bibfnamefont {P.}~\bibnamefont
  {Scheier}}\ and\ \bibinfo {author} {\bibfnamefont {T.~D.}\ \bibnamefont
  {M{\"a}rk}},\ }\bibfield  {title} {\enquote {\bibinfo {title} {Mass-resolved
  argon cluster spectra up to 12000 u (ar$_{300}^+$)},}\ }\href@noop {}
  {\bibfield  {journal} {\bibinfo  {journal} {International Journal of Mass
  Spectrometry and Ion Processes}\ }\textbf {\bibinfo {volume} {76}},\ \bibinfo
  {pages} {R11--R15} (\bibinfo {year} {1987})}\BibitemShut {NoStop}%
\bibitem [{\citenamefont {M{\"a}rk}\ and\ \citenamefont
  {Scheier}(1987)}]{Mark:1987aa}%
  \BibitemOpen
  \bibfield  {author} {\bibinfo {author} {\bibfnamefont {T.~D.}\ \bibnamefont
  {M{\"a}rk}}\ and\ \bibinfo {author} {\bibfnamefont {P.}~\bibnamefont
  {Scheier}},\ }\bibfield  {title} {\enquote {\bibinfo {title} {Production and
  stability of neon cluster ions up to ne$_{90}^+$},}\ }\href {\doibase
  https://doi.org/10.1016/0009-2614(87)80213-0} {\bibfield  {journal} {\bibinfo
   {journal} {Chemical Physics Letters}\ }\textbf {\bibinfo {volume} {137}},\
  \bibinfo {pages} {245--249} (\bibinfo {year} {1987})}\BibitemShut {NoStop}%
\bibitem [{\citenamefont {Gatchell}\ \emph {et~al.}(2018)\citenamefont
  {Gatchell}, \citenamefont {Martini}, \citenamefont {Kranabetter},
  \citenamefont {Rasul},\ and\ \citenamefont {Scheier}}]{Gatchell:2018aa}%
  \BibitemOpen
  \bibfield  {author} {\bibinfo {author} {\bibfnamefont {M.}~\bibnamefont
  {Gatchell}}, \bibinfo {author} {\bibfnamefont {P.}~\bibnamefont {Martini}},
  \bibinfo {author} {\bibfnamefont {L.}~\bibnamefont {Kranabetter}}, \bibinfo
  {author} {\bibfnamefont {B.}~\bibnamefont {Rasul}}, \ and\ \bibinfo {author}
  {\bibfnamefont {P.}~\bibnamefont {Scheier}},\ }\bibfield  {title} {\enquote
  {\bibinfo {title} {Magic sizes of cationic and protonated argon clusters},}\
  }\href@noop {} {\bibfield  {journal} {\bibinfo  {journal} {Physical Review
  A}\ }\textbf {\bibinfo {volume} {98}},\ \bibinfo {pages} {022519} (\bibinfo
  {year} {2018})}\BibitemShut {NoStop}%
\bibitem [{\citenamefont {Gatchell}\ \emph {et~al.}(2019)\citenamefont
  {Gatchell}, \citenamefont {Martini}, \citenamefont {Schiller},\ and\
  \citenamefont {Scheier}}]{Gatchell:2019ac}%
  \BibitemOpen
  \bibfield  {author} {\bibinfo {author} {\bibfnamefont {M.}~\bibnamefont
  {Gatchell}}, \bibinfo {author} {\bibfnamefont {P.}~\bibnamefont {Martini}},
  \bibinfo {author} {\bibfnamefont {A.}~\bibnamefont {Schiller}}, \ and\
  \bibinfo {author} {\bibfnamefont {P.}~\bibnamefont {Scheier}},\ }\bibfield
  {title} {\enquote {\bibinfo {title} {Protonated clusters of neon and
  krypton},}\ }\href {\doibase 10.1007/s13361-019-02329-w} {\bibfield
  {journal} {\bibinfo  {journal} {Journal of The American Society for Mass
  Spectrometry}\ } (\bibinfo {year} {2019}),\
  10.1007/s13361-019-02329-w}\BibitemShut {NoStop}%
\bibitem [{\citenamefont {F{\'a}rnı́k}\ and\ \citenamefont
  {Toennies}(2004)}]{Farnk:2004aa}%
  \BibitemOpen
  \bibfield  {author} {\bibinfo {author} {\bibfnamefont {M.}~\bibnamefont
  {F{\'a}rnı́k}}\ and\ \bibinfo {author} {\bibfnamefont {J.~P.}\ \bibnamefont
  {Toennies}},\ }\bibfield  {title} {\enquote {\bibinfo {title} {Ion-molecule
  reactions in 4he droplets: Flying nano-cryo-reactors},}\ }\bibfield
  {booktitle} {\emph {\bibinfo {booktitle} {The Journal of Chemical Physics}},\
  }\href {\doibase 10.1063/1.1815272} {\bibfield  {journal} {\bibinfo
  {journal} {The Journal of Chemical Physics}\ }\textbf {\bibinfo {volume}
  {122}},\ \bibinfo {pages} {014307} (\bibinfo {year} {2004})}\BibitemShut
  {NoStop}%
\bibitem [{\citenamefont {Toennies}\ and\ \citenamefont
  {Vilesov}(2004)}]{Toennies:2004aa}%
  \BibitemOpen
  \bibfield  {author} {\bibinfo {author} {\bibfnamefont {J.~P.}\ \bibnamefont
  {Toennies}}\ and\ \bibinfo {author} {\bibfnamefont {A.~F.}\ \bibnamefont
  {Vilesov}},\ }\bibfield  {title} {\enquote {\bibinfo {title} {Superfluid
  helium droplets: A uniquely cold nanomatrix for molecules and molecular
  complexes},}\ }\href {\doibase 10.1002/anie.200300611} {\bibfield  {journal}
  {\bibinfo  {journal} {Angewandte Chemie International Edition}\ }\textbf
  {\bibinfo {volume} {43}},\ \bibinfo {pages} {2622--2648} (\bibinfo {year}
  {2004})}\BibitemShut {NoStop}%
\bibitem [{\citenamefont {Laimer}\ \emph {et~al.}(2019)\citenamefont {Laimer},
  \citenamefont {Kranabetter}, \citenamefont {Tiefenthaler}, \citenamefont
  {Albertini}, \citenamefont {Zappa}, \citenamefont {Ellis}, \citenamefont
  {Gatchell},\ and\ \citenamefont {Scheier}}]{Laimer:2019aa}%
  \BibitemOpen
  \bibfield  {author} {\bibinfo {author} {\bibfnamefont {F.}~\bibnamefont
  {Laimer}}, \bibinfo {author} {\bibfnamefont {L.}~\bibnamefont {Kranabetter}},
  \bibinfo {author} {\bibfnamefont {L.}~\bibnamefont {Tiefenthaler}}, \bibinfo
  {author} {\bibfnamefont {S.}~\bibnamefont {Albertini}}, \bibinfo {author}
  {\bibfnamefont {F.}~\bibnamefont {Zappa}}, \bibinfo {author} {\bibfnamefont
  {A.~M.}\ \bibnamefont {Ellis}}, \bibinfo {author} {\bibfnamefont
  {M.}~\bibnamefont {Gatchell}}, \ and\ \bibinfo {author} {\bibfnamefont
  {P.}~\bibnamefont {Scheier}},\ }\bibfield  {title} {\enquote {\bibinfo
  {title} {Highly charged droplets of superfluid helium},}\ }\href {\doibase
  10.1103/PhysRevLett.123.165301} {\bibfield  {journal} {\bibinfo  {journal}
  {Physical Review Letters}\ }\textbf {\bibinfo {volume} {123}},\ \bibinfo
  {pages} {165301--} (\bibinfo {year} {2019})}\BibitemShut {NoStop}%
\bibitem [{\citenamefont {Ralser}\ \emph {et~al.}(2015)\citenamefont {Ralser},
  \citenamefont {Postler}, \citenamefont {Harnisch}, \citenamefont {Ellis},\
  and\ \citenamefont {Scheier}}]{Ralser:2015aa}%
  \BibitemOpen
  \bibfield  {author} {\bibinfo {author} {\bibfnamefont {S.}~\bibnamefont
  {Ralser}}, \bibinfo {author} {\bibfnamefont {J.}~\bibnamefont {Postler}},
  \bibinfo {author} {\bibfnamefont {M.}~\bibnamefont {Harnisch}}, \bibinfo
  {author} {\bibfnamefont {A.~M.}\ \bibnamefont {Ellis}}, \ and\ \bibinfo
  {author} {\bibfnamefont {P.}~\bibnamefont {Scheier}},\ }\bibfield  {title}
  {\enquote {\bibinfo {title} {Extracting cluster distributions from mass
  spectra: Isotopefit},}\ }\href@noop {} {\bibfield  {journal} {\bibinfo
  {journal} {International Journal of Mass Spectrometry}\ }\textbf {\bibinfo
  {volume} {379}},\ \bibinfo {pages} {194--199} (\bibinfo {year}
  {2015})}\BibitemShut {NoStop}%
\bibitem [{\citenamefont {Frisch}\ \emph {et~al.}(2016)\citenamefont {Frisch},
  \citenamefont {Trucks}, \citenamefont {Schlegel}, \citenamefont {Scuseria},
  \citenamefont {Robb}, \citenamefont {Cheeseman}, \citenamefont {Scalmani},
  \citenamefont {Barone}, \citenamefont {Petersson}, \citenamefont {Nakatsuji},
  \citenamefont {Li}, \citenamefont {Caricato}, \citenamefont {Marenich},
  \citenamefont {Bloino}, \citenamefont {Janesko}, \citenamefont {Gomperts},
  \citenamefont {Mennucci}, \citenamefont {Hratchian}, \citenamefont {Ortiz},
  \citenamefont {Izmaylov}, \citenamefont {Sonnenberg}, \citenamefont
  {Williams}, \citenamefont {Ding}, \citenamefont {Lipparini}, \citenamefont
  {Egidi}, \citenamefont {Goings}, \citenamefont {Peng}, \citenamefont
  {Petrone}, \citenamefont {Henderson}, \citenamefont {Ranasinghe},
  \citenamefont {Zakrzewski}, \citenamefont {Gao}, \citenamefont {Rega},
  \citenamefont {Zheng}, \citenamefont {Liang}, \citenamefont {Hada},
  \citenamefont {Ehara}, \citenamefont {Toyota}, \citenamefont {Fukuda},
  \citenamefont {Hasegawa}, \citenamefont {Ishida}, \citenamefont {Nakajima},
  \citenamefont {Honda}, \citenamefont {Kitao}, \citenamefont {Nakai},
  \citenamefont {Vreven}, \citenamefont {Throssell}, \citenamefont
  {Montgomery~Jr.}, \citenamefont {Peralta}, \citenamefont {Ogliaro},
  \citenamefont {Bearpark}, \citenamefont {Heyd}, \citenamefont {Brothers},
  \citenamefont {Kudin}, \citenamefont {Staroverov}, \citenamefont {Keith},
  \citenamefont {Kobayashi}, \citenamefont {Normand}, \citenamefont
  {Raghavachari}, \citenamefont {Rendell}, \citenamefont {Burant},
  \citenamefont {Iyengar}, \citenamefont {Tomasi}, \citenamefont {Cossi},
  \citenamefont {Millam}, \citenamefont {Klene}, \citenamefont {Adamo},
  \citenamefont {Cammi}, \citenamefont {Ochterski}, \citenamefont {Martin},
  \citenamefont {Morokuma}, \citenamefont {Farkas}, \citenamefont {Foresman},\
  and\ \citenamefont {Fox}}]{Frisch:2016aa}%
  \BibitemOpen
  \bibfield  {author} {\bibinfo {author} {\bibfnamefont {M.~J.}\ \bibnamefont
  {Frisch}}, \bibinfo {author} {\bibfnamefont {G.~W.}\ \bibnamefont {Trucks}},
  \bibinfo {author} {\bibfnamefont {H.~B.}\ \bibnamefont {Schlegel}}, \bibinfo
  {author} {\bibfnamefont {G.~E.}\ \bibnamefont {Scuseria}}, \bibinfo {author}
  {\bibfnamefont {M.~A.}\ \bibnamefont {Robb}}, \bibinfo {author}
  {\bibfnamefont {J.~R.}\ \bibnamefont {Cheeseman}}, \bibinfo {author}
  {\bibfnamefont {G.}~\bibnamefont {Scalmani}}, \bibinfo {author}
  {\bibfnamefont {V.}~\bibnamefont {Barone}}, \bibinfo {author} {\bibfnamefont
  {G.~A.}\ \bibnamefont {Petersson}}, \bibinfo {author} {\bibfnamefont
  {H.}~\bibnamefont {Nakatsuji}}, \bibinfo {author} {\bibfnamefont
  {X.}~\bibnamefont {Li}}, \bibinfo {author} {\bibfnamefont {M.}~\bibnamefont
  {Caricato}}, \bibinfo {author} {\bibfnamefont {A.~V.}\ \bibnamefont
  {Marenich}}, \bibinfo {author} {\bibfnamefont {J.}~\bibnamefont {Bloino}},
  \bibinfo {author} {\bibfnamefont {B.~G.}\ \bibnamefont {Janesko}}, \bibinfo
  {author} {\bibfnamefont {R.}~\bibnamefont {Gomperts}}, \bibinfo {author}
  {\bibfnamefont {B.}~\bibnamefont {Mennucci}}, \bibinfo {author}
  {\bibfnamefont {H.~P.}\ \bibnamefont {Hratchian}}, \bibinfo {author}
  {\bibfnamefont {J.~V.}\ \bibnamefont {Ortiz}}, \bibinfo {author}
  {\bibfnamefont {A.~F.}\ \bibnamefont {Izmaylov}}, \bibinfo {author}
  {\bibfnamefont {J.~L.}\ \bibnamefont {Sonnenberg}}, \bibinfo {author}
  {\bibnamefont {Williams}}, \bibinfo {author} {\bibfnamefont {F.}~\bibnamefont
  {Ding}}, \bibinfo {author} {\bibfnamefont {F.}~\bibnamefont {Lipparini}},
  \bibinfo {author} {\bibfnamefont {F.}~\bibnamefont {Egidi}}, \bibinfo
  {author} {\bibfnamefont {J.}~\bibnamefont {Goings}}, \bibinfo {author}
  {\bibfnamefont {B.}~\bibnamefont {Peng}}, \bibinfo {author} {\bibfnamefont
  {A.}~\bibnamefont {Petrone}}, \bibinfo {author} {\bibfnamefont
  {T.}~\bibnamefont {Henderson}}, \bibinfo {author} {\bibfnamefont
  {D.}~\bibnamefont {Ranasinghe}}, \bibinfo {author} {\bibfnamefont {V.~G.}\
  \bibnamefont {Zakrzewski}}, \bibinfo {author} {\bibfnamefont
  {J.}~\bibnamefont {Gao}}, \bibinfo {author} {\bibfnamefont {N.}~\bibnamefont
  {Rega}}, \bibinfo {author} {\bibfnamefont {G.}~\bibnamefont {Zheng}},
  \bibinfo {author} {\bibfnamefont {W.}~\bibnamefont {Liang}}, \bibinfo
  {author} {\bibfnamefont {M.}~\bibnamefont {Hada}}, \bibinfo {author}
  {\bibfnamefont {M.}~\bibnamefont {Ehara}}, \bibinfo {author} {\bibfnamefont
  {K.}~\bibnamefont {Toyota}}, \bibinfo {author} {\bibfnamefont
  {R.}~\bibnamefont {Fukuda}}, \bibinfo {author} {\bibfnamefont
  {J.}~\bibnamefont {Hasegawa}}, \bibinfo {author} {\bibfnamefont
  {M.}~\bibnamefont {Ishida}}, \bibinfo {author} {\bibfnamefont
  {T.}~\bibnamefont {Nakajima}}, \bibinfo {author} {\bibfnamefont
  {Y.}~\bibnamefont {Honda}}, \bibinfo {author} {\bibfnamefont
  {O.}~\bibnamefont {Kitao}}, \bibinfo {author} {\bibfnamefont
  {H.}~\bibnamefont {Nakai}}, \bibinfo {author} {\bibfnamefont
  {T.}~\bibnamefont {Vreven}}, \bibinfo {author} {\bibfnamefont
  {K.}~\bibnamefont {Throssell}}, \bibinfo {author} {\bibfnamefont {J.~A.}\
  \bibnamefont {Montgomery~Jr.}}, \bibinfo {author} {\bibfnamefont {J.~E.}\
  \bibnamefont {Peralta}}, \bibinfo {author} {\bibfnamefont {F.}~\bibnamefont
  {Ogliaro}}, \bibinfo {author} {\bibfnamefont {M.~J.}\ \bibnamefont
  {Bearpark}}, \bibinfo {author} {\bibfnamefont {J.~J.}\ \bibnamefont {Heyd}},
  \bibinfo {author} {\bibfnamefont {E.~N.}\ \bibnamefont {Brothers}}, \bibinfo
  {author} {\bibfnamefont {K.~N.}\ \bibnamefont {Kudin}}, \bibinfo {author}
  {\bibfnamefont {V.~N.}\ \bibnamefont {Staroverov}}, \bibinfo {author}
  {\bibfnamefont {T.~A.}\ \bibnamefont {Keith}}, \bibinfo {author}
  {\bibfnamefont {R.}~\bibnamefont {Kobayashi}}, \bibinfo {author}
  {\bibfnamefont {J.}~\bibnamefont {Normand}}, \bibinfo {author} {\bibfnamefont
  {K.}~\bibnamefont {Raghavachari}}, \bibinfo {author} {\bibfnamefont {A.~P.}\
  \bibnamefont {Rendell}}, \bibinfo {author} {\bibfnamefont {J.~C.}\
  \bibnamefont {Burant}}, \bibinfo {author} {\bibfnamefont {S.~S.}\
  \bibnamefont {Iyengar}}, \bibinfo {author} {\bibfnamefont {J.}~\bibnamefont
  {Tomasi}}, \bibinfo {author} {\bibfnamefont {M.}~\bibnamefont {Cossi}},
  \bibinfo {author} {\bibfnamefont {J.~M.}\ \bibnamefont {Millam}}, \bibinfo
  {author} {\bibfnamefont {M.}~\bibnamefont {Klene}}, \bibinfo {author}
  {\bibfnamefont {C.}~\bibnamefont {Adamo}}, \bibinfo {author} {\bibfnamefont
  {R.}~\bibnamefont {Cammi}}, \bibinfo {author} {\bibfnamefont {J.~W.}\
  \bibnamefont {Ochterski}}, \bibinfo {author} {\bibfnamefont {R.~L.}\
  \bibnamefont {Martin}}, \bibinfo {author} {\bibfnamefont {K.}~\bibnamefont
  {Morokuma}}, \bibinfo {author} {\bibfnamefont {O.}~\bibnamefont {Farkas}},
  \bibinfo {author} {\bibfnamefont {J.~B.}\ \bibnamefont {Foresman}}, \ and\
  \bibinfo {author} {\bibfnamefont {D.~J.}\ \bibnamefont {Fox}},\ }\href@noop
  {} {\enquote {\bibinfo {title} {Gaussian 16 rev. a.03},}\ } (\bibinfo {year}
  {2016})\BibitemShut {NoStop}%
\bibitem [{\citenamefont {Bondi}(1964)}]{Bondi:1964aa}%
  \BibitemOpen
  \bibfield  {author} {\bibinfo {author} {\bibfnamefont {A.}~\bibnamefont
  {Bondi}},\ }\bibfield  {title} {\enquote {\bibinfo {title} {van der waals
  volumes and radii},}\ }\bibfield  {booktitle} {\emph {\bibinfo {booktitle}
  {The Journal of Physical Chemistry}},\ }\href {\doibase 10.1021/j100785a001}
  {\bibfield  {journal} {\bibinfo  {journal} {The Journal of Physical
  Chemistry}\ }\textbf {\bibinfo {volume} {68}},\ \bibinfo {pages} {441--451}
  (\bibinfo {year} {1964})}\BibitemShut {NoStop}%
\bibitem [{\citenamefont {Rowland}\ and\ \citenamefont
  {Taylor}(1996)}]{Rowland:1996aa}%
  \BibitemOpen
  \bibfield  {author} {\bibinfo {author} {\bibfnamefont {R.~S.}\ \bibnamefont
  {Rowland}}\ and\ \bibinfo {author} {\bibfnamefont {R.}~\bibnamefont
  {Taylor}},\ }\bibfield  {title} {\enquote {\bibinfo {title} {Intermolecular
  nonbonded contact distances in organic crystal structures: Comparison with
  distances expected from van der waals radii},}\ }\bibfield  {booktitle}
  {\emph {\bibinfo {booktitle} {The Journal of Physical Chemistry}},\ }\href
  {\doibase 10.1021/jp953141+} {\bibfield  {journal} {\bibinfo  {journal} {The
  Journal of Physical Chemistry}\ }\textbf {\bibinfo {volume} {100}},\ \bibinfo
  {pages} {7384--7391} (\bibinfo {year} {1996})}\BibitemShut {NoStop}%
\bibitem [{\citenamefont {Atkins}(1959)}]{Atkins:1959aa}%
  \BibitemOpen
  \bibfield  {author} {\bibinfo {author} {\bibfnamefont {K.~R.}\ \bibnamefont
  {Atkins}},\ }\bibfield  {title} {\enquote {\bibinfo {title} {Ions in liquid
  helium},}\ }\href {\doibase 10.1103/PhysRev.116.1339} {\bibfield  {journal}
  {\bibinfo  {journal} {Phys. Rev.}\ }\textbf {\bibinfo {volume} {116}},\
  \bibinfo {pages} {1339--1343} (\bibinfo {year} {1959})}\BibitemShut {NoStop}%
\bibitem [{\citenamefont {Buzzacchi}, \citenamefont {Galli},\ and\
  \citenamefont {Reatto}(2001)}]{Buzzacchi:2001aa}%
  \BibitemOpen
  \bibfield  {author} {\bibinfo {author} {\bibfnamefont {M.}~\bibnamefont
  {Buzzacchi}}, \bibinfo {author} {\bibfnamefont {D.~E.}\ \bibnamefont
  {Galli}}, \ and\ \bibinfo {author} {\bibfnamefont {L.}~\bibnamefont
  {Reatto}},\ }\bibfield  {title} {\enquote {\bibinfo {title} {Alkali ions in
  superfluid ${}^{4}\mathrm{He}$ and structure of the snowball},}\ }\href
  {\doibase 10.1103/PhysRevB.64.094512} {\bibfield  {journal} {\bibinfo
  {journal} {Phys. Rev. B}\ }\textbf {\bibinfo {volume} {64}},\ \bibinfo
  {pages} {094512} (\bibinfo {year} {2001})}\BibitemShut {NoStop}%
\bibitem [{\citenamefont {M{\"u}ller}, \citenamefont {Mudrich},\ and\
  \citenamefont {Stienkemeier}(2009)}]{Muller:2009ab}%
  \BibitemOpen
  \bibfield  {author} {\bibinfo {author} {\bibfnamefont {S.}~\bibnamefont
  {M{\"u}ller}}, \bibinfo {author} {\bibfnamefont {M.}~\bibnamefont {Mudrich}},
  \ and\ \bibinfo {author} {\bibfnamefont {F.}~\bibnamefont {Stienkemeier}},\
  }\bibfield  {title} {\enquote {\bibinfo {title} {Alkali-helium snowball
  complexes formed on helium nanodroplets},}\ }\href {\doibase
  10.1063/1.3180819} {\bibfield  {journal} {\bibinfo  {journal} {The Journal of
  Chemical Physics}\ }\textbf {\bibinfo {volume} {131}},\ \bibinfo {pages}
  {044319} (\bibinfo {year} {2009})}\BibitemShut {NoStop}%
\bibitem [{\citenamefont {Leidlmair}\ \emph {et~al.}(2012)\citenamefont
  {Leidlmair}, \citenamefont {Wang}, \citenamefont {Bartl}, \citenamefont
  {Sch\"obel}, \citenamefont {Denifl}, \citenamefont {Probst}, \citenamefont
  {Alcam\'i}, \citenamefont {Mart\'in}, \citenamefont {Zettergren},
  \citenamefont {Hansen}, \citenamefont {Echt},\ and\ \citenamefont
  {Scheier}}]{PhysRevLett.108.076101}%
  \BibitemOpen
  \bibfield  {author} {\bibinfo {author} {\bibfnamefont {C.}~\bibnamefont
  {Leidlmair}}, \bibinfo {author} {\bibfnamefont {Y.}~\bibnamefont {Wang}},
  \bibinfo {author} {\bibfnamefont {P.}~\bibnamefont {Bartl}}, \bibinfo
  {author} {\bibfnamefont {H.}~\bibnamefont {Sch\"obel}}, \bibinfo {author}
  {\bibfnamefont {S.}~\bibnamefont {Denifl}}, \bibinfo {author} {\bibfnamefont
  {M.}~\bibnamefont {Probst}}, \bibinfo {author} {\bibfnamefont
  {M.}~\bibnamefont {Alcam\'i}}, \bibinfo {author} {\bibfnamefont
  {F.}~\bibnamefont {Mart\'in}}, \bibinfo {author} {\bibfnamefont
  {H.}~\bibnamefont {Zettergren}}, \bibinfo {author} {\bibfnamefont
  {K.}~\bibnamefont {Hansen}}, \bibinfo {author} {\bibfnamefont
  {O.}~\bibnamefont {Echt}}, \ and\ \bibinfo {author} {\bibfnamefont
  {P.}~\bibnamefont {Scheier}},\ }\bibfield  {title} {\enquote {\bibinfo
  {title} {Structures, energetics, and dynamics of helium adsorbed on isolated
  fullerene ions},}\ }\href {\doibase 10.1103/PhysRevLett.108.076101}
  {\bibfield  {journal} {\bibinfo  {journal} {Physical Review Letters}\
  }\textbf {\bibinfo {volume} {108}},\ \bibinfo {pages} {076101} (\bibinfo
  {year} {2012})}\BibitemShut {NoStop}%
\bibitem [{\citenamefont {Calvo}(2015)}]{Calvo:2015aa}%
  \BibitemOpen
  \bibfield  {author} {\bibinfo {author} {\bibfnamefont {F.}~\bibnamefont
  {Calvo}},\ }\bibfield  {title} {\enquote {\bibinfo {title} {Coating
  polycyclic aromatic hydrocarbon cations with helium clusters: Snowballs and
  slush},}\ }\bibfield  {booktitle} {\emph {\bibinfo {booktitle} {The Journal
  of Physical Chemistry A}},\ }\href {\doibase 10.1021/jp510799h} {\bibfield
  {journal} {\bibinfo  {journal} {The Journal of Physical Chemistry A}\
  }\textbf {\bibinfo {volume} {119}},\ \bibinfo {pages} {5959--5970} (\bibinfo
  {year} {2015})}\BibitemShut {NoStop}%
\bibitem [{\citenamefont {Kuhn}\ \emph {et~al.}(2016)\citenamefont {Kuhn},
  \citenamefont {Renzler}, \citenamefont {Postler}, \citenamefont {Ralser},
  \citenamefont {Spieler}, \citenamefont {Simpson}, \citenamefont {Linnartz},
  \citenamefont {Tielens}, \citenamefont {Cami}, \citenamefont {Mauracher},
  \citenamefont {Wang}, \citenamefont {Alcam{\'\i}}, \citenamefont
  {Mart{\'\i}n}, \citenamefont {Beyer}, \citenamefont {Wester}, \citenamefont
  {Lindinger},\ and\ \citenamefont {Scheier}}]{Kuhn:2016aa}%
  \BibitemOpen
  \bibfield  {author} {\bibinfo {author} {\bibfnamefont {M.}~\bibnamefont
  {Kuhn}}, \bibinfo {author} {\bibfnamefont {M.}~\bibnamefont {Renzler}},
  \bibinfo {author} {\bibfnamefont {J.}~\bibnamefont {Postler}}, \bibinfo
  {author} {\bibfnamefont {S.}~\bibnamefont {Ralser}}, \bibinfo {author}
  {\bibfnamefont {S.}~\bibnamefont {Spieler}}, \bibinfo {author} {\bibfnamefont
  {M.}~\bibnamefont {Simpson}}, \bibinfo {author} {\bibfnamefont
  {H.}~\bibnamefont {Linnartz}}, \bibinfo {author} {\bibfnamefont {A.~G.
  G.~M.}\ \bibnamefont {Tielens}}, \bibinfo {author} {\bibfnamefont
  {J.}~\bibnamefont {Cami}}, \bibinfo {author} {\bibfnamefont {A.}~\bibnamefont
  {Mauracher}}, \bibinfo {author} {\bibfnamefont {Y.}~\bibnamefont {Wang}},
  \bibinfo {author} {\bibfnamefont {M.}~\bibnamefont {Alcam{\'\i}}}, \bibinfo
  {author} {\bibfnamefont {F.}~\bibnamefont {Mart{\'\i}n}}, \bibinfo {author}
  {\bibfnamefont {M.~K.}\ \bibnamefont {Beyer}}, \bibinfo {author}
  {\bibfnamefont {R.}~\bibnamefont {Wester}}, \bibinfo {author} {\bibfnamefont
  {A.}~\bibnamefont {Lindinger}}, \ and\ \bibinfo {author} {\bibfnamefont
  {P.}~\bibnamefont {Scheier}},\ }\bibfield  {title} {\enquote {\bibinfo
  {title} {Atomically resolved phase transition of fullerene cations solvated
  in helium droplets},}\ }\href@noop {} {\bibfield  {journal} {\bibinfo
  {journal} {Nature Communications}\ }\textbf {\bibinfo {volume} {7}},\
  \bibinfo {pages} {13550} (\bibinfo {year} {2016})}\BibitemShut {NoStop}%
\bibitem [{\citenamefont {Rastogi}\ \emph {et~al.}(2018)\citenamefont
  {Rastogi}, \citenamefont {Leidlmair}, \citenamefont {An~der Lan},
  \citenamefont {Ortiz~de Z{\'a}rate}, \citenamefont {P{\'e}rez~de Tudela},
  \citenamefont {Bartolomei}, \citenamefont {Hern{\'a}ndez}, \citenamefont
  {Campos-Mart{\'\i}nez}, \citenamefont {Gonz{\'a}lez-Lezana}, \citenamefont
  {Hern{\'a}ndez-Rojas}, \citenamefont {Bret{\'o}n}, \citenamefont {Scheier},\
  and\ \citenamefont {Gatchell}}]{Rastogi:2018aa}%
  \BibitemOpen
  \bibfield  {author} {\bibinfo {author} {\bibfnamefont {M.}~\bibnamefont
  {Rastogi}}, \bibinfo {author} {\bibfnamefont {C.}~\bibnamefont {Leidlmair}},
  \bibinfo {author} {\bibfnamefont {L.}~\bibnamefont {An~der Lan}}, \bibinfo
  {author} {\bibfnamefont {J.}~\bibnamefont {Ortiz~de Z{\'a}rate}}, \bibinfo
  {author} {\bibfnamefont {R.}~\bibnamefont {P{\'e}rez~de Tudela}}, \bibinfo
  {author} {\bibfnamefont {M.}~\bibnamefont {Bartolomei}}, \bibinfo {author}
  {\bibfnamefont {M.~I.}\ \bibnamefont {Hern{\'a}ndez}}, \bibinfo {author}
  {\bibfnamefont {J.}~\bibnamefont {Campos-Mart{\'\i}nez}}, \bibinfo {author}
  {\bibfnamefont {T.}~\bibnamefont {Gonz{\'a}lez-Lezana}}, \bibinfo {author}
  {\bibfnamefont {J.}~\bibnamefont {Hern{\'a}ndez-Rojas}}, \bibinfo {author}
  {\bibfnamefont {J.}~\bibnamefont {Bret{\'o}n}}, \bibinfo {author}
  {\bibfnamefont {P.}~\bibnamefont {Scheier}}, \ and\ \bibinfo {author}
  {\bibfnamefont {M.}~\bibnamefont {Gatchell}},\ }\bibfield  {title} {\enquote
  {\bibinfo {title} {Lithium ions solvated in helium},}\ }\href {\doibase
  10.1039/C8CP04522D} {\bibfield  {journal} {\bibinfo  {journal} {Phys. Chem.
  Chem. Phys.}\ }\textbf {\bibinfo {volume} {20}},\ \bibinfo {pages}
  {25569--25576} (\bibinfo {year} {2018})}\BibitemShut {NoStop}%
\end{thebibliography}%

\end{document}